\documentclass[12pt,a4paper]{article}

\usepackage[utf8]{inputenc}
\usepackage[T1]{fontenc}
\usepackage{lmodern}
\usepackage{amsmath,amssymb,amsthm}
\usepackage{graphicx}
\usepackage{caption}
\usepackage{subcaption}
\usepackage{booktabs}
\usepackage{tabularx}
\usepackage{braket}
\usepackage{dsfont}
\usepackage{float}

\usepackage{hyperref}
\hypersetup{
	colorlinks=true,
	linkcolor=blue,
	citecolor=blue,
	urlcolor=blue,
}

\usepackage{geometry}
\usepackage{fancyhdr}
\newtheorem{remark}{Remark}
\newtheorem{example}{Example}

\title{%
	\vspace{2cm}
	Packaged Quantum States for Quantum Simulation of Lattice Gauge Theories
}

\author{
	Rongchao Ma \\
	\textit{Department of Physics, University of Alberta, Edmonton, Canada}\\
}

\date{\today}

\begin{document}
	
\maketitle

\begin{abstract}
We develop a mathematical framework for the quantum simulation of lattice gauge theories using gauge-invariant packaged quantum states \cite{Ma2017,Ma2025}.
In this formalism, every single excitation transforms as a complete \textbf{irreducible representation (irrep)} of the local gauge group, preventing any appearance of fractional or partial \textbf{internal quantum numbers (IQNs)}.
Multi-particle excitations can form nontrivial packaged entangled states that are also gauge invariant, thereby forbidding partial or fractional IQNs.
In other words, all IQNs of such packaged entangled states remain inseparably entangled.
This ``packaging principle'' ensures that physical states remain confined to the correct gauge sector and excludes partial charges or colors, even when multiple excitations are entangled.

We illustrate this approach for $\mathrm{U}(1)$, $\mathrm{SU}(2)$, and $\mathrm{SU}(3)$ lattice gauge theories, discussing explicit constructions, Trotterized Hamiltonian evolution, and gauge-invariant measurements on a quantum simulator.
We also outline how packaged states can mitigate gauge-violating errors and serve as natural building blocks for gauge-invariant coding schemes, while noting that standard quantum error correction is still required against typical local noise that respects gauge symmetry.
\end{abstract}

\tableofcontents

\section{Introduction}

\textbf{Packaged entangled states} \cite{Ma2017,Ma2025} are multi-particle quantum states in which all \textbf{internal quantum numbers (IQNs)} are inseparably entangled and thus ``packaged'' together, prohibiting any partial or fractional charges or colors. 
For a single excitation, this packaging means that each particle operator carries a complete \textbf{irreducible representation (irrep)} \cite{Weyl1925,Wigner1939} of the local gauge group. 
Although single-particle excitations are not entangled by themselves, multi-particle superpositions within a fixed gauge sector can produce highly nontrivial packaged entangled states.

For example, consider the packaged entangled electron–positron pair \cite{Ma2025},
\[
\frac{1}{\sqrt{2}} 
\bigl[
\hat{a}_{e^-}^\dagger(\mathbf{p}_1)\,\hat{b}_{e^+}^\dagger(\mathbf{p}_2)
\;+\;
\hat{b}_{e^+}^\dagger(\mathbf{p}_1)\,\hat{a}_{e^-}^\dagger(\mathbf{p}_2)
\bigr]
\lvert 0\rangle,
\]
where $\hat{a}_{e^-}^\dagger$ creates an electron $(Q=-e)$ and $\hat{b}_{e^+}^\dagger$ creates a positron $(Q=+e)$. 
The relevant IQNs are electric charge $(Q)$, lepton number $(L)$, flavor $(L_e)$, weak isospin $(T_3)$, and weak hypercharge $(Y)$. 
Within this electron–positron pair, these IQNs $\{Q, L, L_e, T_3, Y\}$ are inseparably entangled; one cannot entangle only the charge $Q$ and leave the other quantum numbers unentangled.

In a gauge theory context \cite{PeskinSchroeder,WeinbergBook}, this principle is expressed via local gauge invariance \cite{Feynman1949,Yang1954,Utiyama1956,Weinberg1967} and superselection rules \cite{WWW1952,DHR1,DHR2,StreaterWightman}: 
physical processes and states cannot coherently mix charge or color sectors, nor can they break Gauss’s law constraints. 
Recent advances in quantum computing and quantum simulation \cite{NielsenChuang} have spurred major efforts toward realizing \textbf{lattice gauge theories (LGTs)} \cite{Wilson1974,KogutSusskind1975} on quantum hardware \cite{Zohar2016,Rothe2012,Banerjee2013,Hauke2013,Tagliacozzo2013,KeeverLubasch2023,KeeverLubasch2024}, driven by:
(1)~the desire to study nonperturbative phenomena in strongly coupled regimes (such as confinement or hadron spectroscopy), and 
(2)~the potential of gauge theories for robust quantum-computing architectures.

Maintaining gauge invariance on real quantum hardware poses a major challenge due to imperfections. Hence, a significant line of research focuses on designing robust encodings to keep the system in its physical, gauge-invariant subspace. 
Here, we show how packaged states address this challenge:
each local excitation is realized as a full gauge-group irrep, so Gauss’s law is automatically satisfied, preventing partial charges or colors from appearing.
In multi-particle configurations, nontrivial packaged entanglement remains gauge invariant, ensuring no fractional or partial IQNs:
forming integer flux loops in $\mathrm{U}(1)$ \cite{Polyakov1975,Luscher1980}, color-confining flux tubes in $\mathrm{SU}(2)$ \cite{Bali1995,Bali2001,Takahashi2002}, or color-singlet hadrons in $\mathrm{SU}(3)$ QCD-like models.
Moreover, we demonstrate how this packaging principle can mitigate unphysical gauge-violating processes, by making them readily detectable as leakage outside the physical subspace - though standard quantum error-correction is still required to combat typical hardware noise that does not break the gauge.

In what follows,
\begin{enumerate}
	\item We develop the general formalism for lattice gauge theories and show how packaging naturally arises from local gauge invariance and superselection (\S\ref{sec:GeneralFormalism}).
	
	\item We discuss how multi-particle packaging leads to nontrivial entangled states, with potential applications in gauge-invariant codes and error mitigation (\S\ref{sec:ApplicationsOfPQS0}).
		
	\item We illustrate these ideas explicitly for $\mathrm{U}(1)$, $\mathrm{SU}(2)$, and $\mathrm{SU}(3)$ gauge theories (\S\ref{sec:U1Gauge}--\ref{sec:SU3Gauge}), detailing Trotterized Hamiltonian schemes \cite{Trotter1959,Suzuki1976,Suzuki1990,Zohar2011,Banerjee2012} and gauge-invariant measurements.
\end{enumerate}

Our aim is a self-consistent strategy for quantum simulation of lattice gauge theories that ``builds in'' gauge invariance from the start---reducing the need for large penalty terms or gauge fixing whenever hardware constraints permit.

\section{General Formalism for Quantum Simulations of LGTs}
\label{sec:GeneralFormalism}

In this section, we develop a broad mathematical framework for quantum simulation of \textbf{lattice gauge theories (LGTs)}, with particular focus on the concept of packaged states, i.e., gauge-invariant excitations built from irreps of the local gauge group. 
We show how local gauge invariance, superselection rules, and irreducible representations (irreps) naturally give rise to ``packaged’’ excitations in both single-particle and multi-particle sectors.
For single-particle excitations, each quantum of matter or flux transforms in a single irreps of the gauge group; for multi-particle states, superpositions of different excitations can form entangled, yet still gauge-invariant, configurations.  

We begin with a brief overview of lattice gauge theories and the essential idea of local gauge symmetry.
Next, we define the physical Hilbert space via Gauss’s law constraints and show how ``packaging'' ensures gauge invariance.
We then discuss truncated link models, such as \textbf{quantum link models (QLMs)}, which offer a route to finite-dimensional implementations on quantum hardware.
Finally, we outline how to carry out both analog and digital quantum simulations, including Trotter-based approaches, error considerations, and the preparation/measurement of packaged states.

\subsection{Formulation of Lattice Gauge Theories}

A lattice gauge theory \cite{Wilson1974,KogutSusskind1975} replaces continuous spacetime with a discrete grid (sites connected by links) and assigns:
\begin{enumerate}
	\item Matter fields (e.g.\ quarks) to lattice \emph{sites},
	  
	\item Gauge fields (e.g.\ gluons) to \emph{links} connecting the sites.
\end{enumerate}

In continuum gauge theory, gauge transformations shift the gauge fields $A_\mu$ without affecting physical observables.  On the lattice, each link $\ell$ carries a group element $U_\ell\in G$, where $G$ might be $\mathrm{U}(1)$, $\mathrm{SU}(2)$, $\mathrm{SU}(3)$, or a different gauge group $\mathrm{SU}(N)$.  Local gauge invariance demands that at each site $x$, one can apply an independent element $g_x \in G$ that simultaneously transforms all links emanating from $x$ (and also transforms the matter field at $x$), leaving the physics unchanged.

\paragraph{Discrete Lattice Setup.}
\begin{itemize}
	\item Let $\Lambda$ be a $d$-dimensional lattice with sites labeled by $x\in \Lambda$.  
	
	\item Each \emph{site} $x$ can host matter fields $\hat{\psi}_x$.  
	
	\item Each \emph{link} $\ell = \langle x,y\rangle$ carries a gauge field operator $\hat{U}_\ell$.
\end{itemize}

For simplicity, one often considers a hypercubic lattice, though all arguments extend to other geometries.

\subsection{Hamiltonian Formulation and Gauss’s Law}
\label{sec:HamiltonianGaussLaw}

To simulate the real-time dynamics of a gauge theory, it is usually more convenient to adopt the Hamiltonian formulation \cite{KogutSusskind1975,Chandrasekharan1997,Wiese2014}.  In a typical Hamiltonian LGT for group $G$, one has terms for electric energy, magnetic (plaquette) energy, and possible matter-gauge couplings:
\begin{equation}\label{GaugeTheoryHamiltonians}
	\hat{H}
	\;=\;
	\underbrace{\sum_{\ell}\frac{g^2}{2}\,\hat{E}_\ell^2}_\text{electric energy}
	\;+\;
	\underbrace{\sum_{\Box}
		\frac{1}{2g^2}
		\bigl(\hat{U}_\Box + \hat{U}_\Box^\dagger\bigr)}_\text{magnetic/plaquette}
	\;+\;
	\underbrace{\sum_{x,\ell}\bigl[
		\hat{\bar{\psi}}_{x}(\gamma^\mu)\,\hat{U}_\ell\,\hat{\psi}_{x+\ell} + \dots
		\bigr]}_\text{matter-kinetic}
	\;+\;\dots
\end{equation}
Here, $\hat{U}_\ell$ creates or annihilates gauge flux on link $\ell$, and $\hat{E}_\ell$ is the ``electric field'' operator.  These operators must obey the \emph{local gauge constraints}, typically referred to as \emph{Gauss’s law}:
\begin{equation}\label{GaussLaw}
	\hat{G}_x\,\ket{\Psi_{\text{phys}}} = \ket{\Psi_{\text{phys}}}, 
	\quad
	\forall x.
\end{equation}
In the Abelian case ($G = \mathrm{U}(1)$), this amounts to 
$\sum_{\ell \text{ emanating from } x}\hat{E}_\ell^{} \;=\; \hat{\rho}_x$,
where $\hat{\rho}_x$ is the matter charge at site $x$.  For non-Abelian groups, one similarly requires the sum of color flux to match the local color charge, e.g.
$
\sum_{\ell} \hat{E}_\ell^a = \hat{T}^a_x
$
with Lie algebra indices $a$.

All physical states must reside in the \emph{gauge-invariant subspace} $\mathcal{H}_{\mathrm{phys}}\subset \mathcal{H}$.  This means physical states cannot carry ``fractional'' gauge quantum numbers; they must form total color singlets (or, more generally, an invariant irrep combination with the matter fields).  As we shall see, this enforced structure underlies the notion of \emph{packaged excitations}.

\subsection{Physical Hilbert Space and Packaged Excitations}
\label{sec:PhysicalHilbertSpace}

In the Hamiltonian approach, the total Hilbert space naively factorizes over all links and sites:
\[
\mathcal{H} 
\;=\; 
\bigotimes_{\ell\in\text{links}} \mathcal{H}_\ell^{(g)} 
\;\otimes\;
\bigotimes_{x\in\text{sites}} \mathcal{H}_x^{(m)}.
\]
However, not all states in this huge product are physically allowed; one must impose Gauss’s law at each site.  
\begin{itemize}
	\item Matter Hilbert space $\mathcal{H}_x^{(m)}$:  This space can be fermionic (quarks), bosonic, or otherwise, but transforms under some representation of the local gauge group $G$.  
	
	\item Gauge Hilbert space $\mathcal{H}_\ell^{(g)}$:  Each link’s gauge field belongs to an (often infinite-dimensional) representation of $G$.  For $\mathrm{U}(1)$, one can think of an integer flux basis; for $\mathrm{SU}(2)$, a sum over spin-$j$ representations; etc.
\end{itemize}

Gauss’s law picks out the subspace
\[
\mathcal{H}_{\mathrm{phys}}
\;=\;
\Bigl\{
\ket{\Psi}\in \mathcal{H}
\;\Big\vert\;
\hat{G}_x \ket{\Psi} = \ket{\Psi},\;\forall x
\Bigr\}.
\]
The essence is: \emph{physical excitations are always complete ``packages'' of gauge quantum numbers that sum to a singlet at every site.}

\paragraph{Single-Particle Packaging.}
A single excitation (e.g., one quark) at site $x$ transforms in a fundamental irrep of $G$.  Because of Gauss’s law, the rest of the system (the link flux emanating from $x$) must precisely ``compensate'' that irrep so that the total color or charge at $x$ remains neutral.  This is what it means for the state to be packaged:  the matter excitation \emph{and} the local flux excitations form a gauge-invariant block together.

\paragraph{Multi-Particle Entangled Packaging.}
When multiple quarks or gluons appear, they can form composite color singlets (like hadrons in QCD) or flux loops, again ensuring net gauge invariance.  In the quantum simulator, these \emph{multi-particle packaged states} can exhibit non-trivial entanglement while still obeying local constraints. Superpositions of hadronic excitations, flux tubes \cite{Bali1995,Bali2001,Takahashi2002,Zohar2011}, or more complicated color structures remain gauge-invariant by virtue of packaging the irreps consistently.

\subsection{Truncating the Gauge Fields: Quantum Link Models}
\label{sec:QLMTruncation}

In standard lattice gauge theories, the link Hilbert space is infinite-dimensional: for $\mathrm{U}(1)$, one can have arbitrarily large integer flux, and for non-Abelian groups like $\mathrm{SU}(2)$ or $\mathrm{SU}(3)$, there is an infinite tower of irreps.  On quantum hardware, however, each link must be encoded in a finite-dimensional system.

A widely employed solution is the \emph{quantum link model} (QLM) \cite{Zohar2016,Chandrasekharan1997,Wiese2014,Sala2018}, which truncates the gauge fields by replacing each link’s infinite-dimensional Hilbert space with a finite representation of the gauge group:

\begin{itemize}
	\item $\mathrm{U}(1)$ QLM:  The integer-valued fluxes $E_\ell \in \mathbb{Z}$ are restricted to a range $-S,\dots,+S$.  One may then interpret $\hat{E}_\ell$ as the spin-$z$ operator $S^z_\ell$ of a spin-$S$ system, while the link operators $\hat{U}_\ell$ become the raising/lowering operators $S^+_\ell$ and $S^-_\ell$.
	
	\item $\mathrm{SU}(2)$ QLM:  Each link is placed in a spin-$S$ representation of $\mathrm{SU}(2)$.  This confines color flux states to the $\dim(2S+1)$ states of that representation, with the three generators of $\mathrm{SU}(2)$ acting within this finite subspace.
	
	\item $\mathrm{SU}(3)$ QLM:  Each link carries a finite-dimensional $\mathrm{SU}(3)$ representation, labeled for instance by Dynkin indices $[p,q]$.  This representation has dimension $\tfrac{1}{2}(p+1)(q+1)(p+q+2)$, limiting the allowed color flux states to that subspace.  The eight generators $\hat{E}^a_\ell\,(a=1,\ldots,8)$ act within this truncated Hilbert space, and the link operators $\hat{U}_\ell$ serve as ladder operators connecting different flux states.
\end{itemize}

A crucial requirement is that the truncated model must be still gauge invariant, i.e., in  all \emph{truncated} operators still \emph{commute} with (or transform consistently under) the local gauge transformations $\hat{G}_x$. In other words, 
\[
[\,\hat{G}_x,\;\hat{U}_\ell\,] \;=\; 0
\quad
\text{and}
\quad
[\,\hat{G}_x,\;\hat{E}_\ell^a\,] \;=\; 0
\]
must hold within the truncated subspace, thus preserving Gauss's law. The quantum link model construction ensures this by defining $\hat{S}^a_\ell$ and $\hat{U}_\ell$ in a manner consistent with the gauge group's algebra, albeit up to finite cutoffs.

Moreover, because each link is now \emph{finite-dimensional}, it can be directly encoded in qubits or qudits.  For instance, if $2S+1$ is a power of 2, then a spin-$S$ link can be mapped onto $\log_2(2S+1)$ qubits.  Alternatively, one could employ a single qudit of dimension $d = 2S+1$.

Although the QLM only approximates the original gauge theory, increasing the representation’s dimension (by enlarging $S$ or the $\mathrm{SU}(3)$ Dynkin labels $[p,q]$) systematically improves the fidelity to the full theory, making quantum link models a valuable framework for hardware-friendly lattice gauge theory simulations.

\subsection{Matter Fields and Their Packaging}
\label{sec:MatterFieldsPackaging}

Many lattice gauge theories of interest include \emph{matter fields} - for instance, quarks in QCD - placed on the lattice sites \cite{WilsonBook1977,Banks1976,Susskind1977}. These matter fields transform under some representation of the gauge group $G$, which ensures that the total color or charge at each site, plus the flux on the adjoining links, must form a color/charge singlet.  This is the essence of the \emph{packaging principle}, preventing unphysical fractional charges and partial IQNs entanglements.

\paragraph{Fermionic Matter (e.g.\ quarks).}
In high-energy physics, matter fields are typically \emph{fermionic}, such as quarks that transform under the fundamental representation $\mathbf{3}$ of $\mathrm{SU}(3)$.  On the lattice, each site $x$ then hosts a set of fermionic creation/annihilation operators,
\[
\hat{\psi}_{x,\alpha}, 
\quad 
\hat{\psi}_{x,\alpha}^\dagger,
\quad 
\alpha \in \{\text{color},\text{spin},\dots\},
\]
satisfying the anticommutation relations
\begin{equation}
	\label{AntiCommutationRelations}
	\{\hat{\psi}_{x,\alpha},\, \hat{\psi}_{x',\beta}^\dagger \}
	\;=\; \delta_{x,x'}\,\delta_{\alpha,\beta},
	\quad
	\{\hat{\psi}_{x,\alpha},\, \hat{\psi}_{x',\beta}\}
	\;=\; 0.
\end{equation}
Each quark excitation carries a ``color'' index $\alpha$ in $\mathbf{3}$ (or $\overline{\mathbf{3}}$) of $\mathrm{SU}(3)$, so the on-site fermionic Hilbert space grows by one two-dimensional occupation subspace per color/spin mode (occupied vs.\ unoccupied).

Enforcing Gauss’s law at site $x$ means the net color flux on outgoing links plus the local quark color must sum to a total singlet.  Concretely, if a quark is present, it contributes a fundamental color $\mathbf{3}$ or $\overline{\mathbf{3}}$; the link fluxes must then compensate to form an overall color-neutral combination.

\paragraph{Bosonic Matter.}
In some settings - such as condensed-matter analogs of gauge theories or certain simplified toy models - one introduces \emph{bosonic} matter fields at each site $x$.  These are created/annihilated by
\[
\hat{\phi}_{x,\alpha}, 
\quad 
\hat{\phi}_{x,\alpha}^\dagger,
\]
with standard commutation relations
\begin{equation}
	\label{CommutationRelations}
	[\hat{\phi}_{x,\alpha},\, \hat{\phi}_{x',\beta}^\dagger]
	= \delta_{x,x'}\,\delta_{\alpha,\beta},
	\quad
	[\hat{\phi}_{x,\alpha},\, \hat{\phi}_{x',\beta}]
	= 0.
\end{equation}
Each bosonic excitation likewise transforms under a representation $\mathbf{d}$ of the gauge group (e.g.\ $\mathbf{2}$ for $\mathrm{SU}(2)$, $\mathbf{3}$ for $\mathrm{SU}(3)$), so ``packaging'' ensures the total color plus flux remains in a singlet.  Unlike fermions, multiple bosons can occupy the same mode, expanding the local Fock space dimension accordingly.

\paragraph{Hybrid Fock Space and Local Constraints.}
Regardless of whether matter is fermionic or bosonic, the local site Hilbert space $\mathcal{R}_x$ is built from:
\begin{itemize}
	\item The gauge representation space for color or charge indices,
	\item The appropriate (anti)commutation relations (fermionic vs.\ bosonic),
	\item Additional quantum numbers such as spin or flavor, if present.
\end{itemize}
Gauss’s law enforces that the on-site matter representation and the link fluxes together form a gauge singlet: 
\[
\hat{G}_x\,\ket{\Psi} \;=\; \ket{\Psi}.
\]
Thus, \emph{a single matter excitation necessarily comes with the appropriate flux to remain gauge-invariant}, and multi-particle states must sum to zero net color or charge at every site.

\paragraph{Practical Implications.}
\begin{enumerate}
	\item \emph{QCD-like Simulations ($\mathrm{SU}(3)$ with Quarks).} Each site can host up to three colors of fermions per spin/flavor mode, with Jordan--Wigner or Bravyi--Kitaev mappings typically needed to encode the fermionic operators into qubits or qudits.  The packaging principle ensures that no unphysical partial color excitations appear.
	\item \emph{Bosons in Optical Lattices.} Ultracold-atom experiments can realize bosonic matter fields with $\mathrm{SU}(N)$ color degrees of freedom.  Sites may be occupied by multiple bosons (up to a truncation), but Gauss’s law still demands overall color neutrality with the link fluxes.
\end{enumerate}
Once these constraints are encoded, the time-evolution (e.g.\ Trotterized steps), measurements (Wilson loops, flux densities, etc.), and error considerations (see Sec.~\ref{sec:ErrorChannelsGaugeConstraints}) proceed analogously to the pure gauge or simpler $\mathrm{U}(1)$ cases.

\paragraph{Gauge Invariance Enforcement}
In practice, one enforces Gauss’s law either by adding large penalty terms $\lambda\,(\hat{G}_x - I)^2$ to the Hamiltonian or by choosing an encoding that spans only the physical, gauge-invariant subspace from the start.  In either approach, the packaging principle ensures that matter excitations remain color- or charge-neutral once combined with the appropriate flux, prohibiting the appearance of unphysical fractional charges.

\subsection{Analog vs.\ Digital Quantum Simulation Strategies}
\label{sec:AnalogDigital}

Once the Hilbert space and truncations are specified, one can simulate the real-time evolution or ground-state properties of the LGT on a quantum device.

\subsubsection{Analog Quantum Simulation}
In analog approaches \cite{Banerjee2013}, common in cold-atom and trapped-ion platforms, one engineers a physical system whose \emph{native} Hamiltonian matches (or closely approximates) the target LGT Hamiltonian Eq. \eqref{GaugeTheoryHamiltonians}.  
Gauge invariance is enforced by carefully tuned interactions and energy penalties that suppress gauge-violating processes.  The simulator then evolves ``naturally'' under this gauge-invariant Hamiltonian, enabling the direct measurement of correlation functions, Wilson loops, and other observables.

\subsubsection{Digital Quantum Simulation}
In digital approaches \cite{Jordan2012}, one implements the time-evolution operator 
\[
e^{-iH\Delta t}
\]
via a Trotter-Suzuki decomposition:
\begin{equation}\label{Eq:TrotterSuzukiDecomposition}
	e^{-iH \Delta t} 
	\;\approx\;
	\prod_{\alpha} \; e^{-i H_{\alpha}\,\Delta t},
\end{equation}
where each $H_{\alpha}$ is a simpler (often local) piece of the Hamiltonian (e.g., the electric part $H_E$ and the magnetic part $H_B$).  
This product is then realized as a quantum circuit of elementary gates acting on qubits or qudits.  By iterating many small time steps, one simulates evolution over a longer duration $T$.

There are two broad strategies:
\begin{itemize}
	\item \textbf{Gauge Invariance:} In an ideal circuit, each Trotter step respects Gauss’s law by construction.  In practice, small Trotter errors or hardware noise can introduce gauge violations, which one mitigates via penalty terms or gauge-constraint measurements (see Sec.~\ref{sec:ErrorChannelsGaugeConstraints}).
	
	\item \textbf{Variational Approaches:} One may also use a variational quantum eigensolver (VQE) \cite{Peruzzo2014} to find low-energy gauge-invariant states.  Here, a parameterized circuit prepares trial states, and the variational algorithm measures the energy while enforcing (or penalizing deviations from) gauge symmetry.
\end{itemize}

Recent work \cite{KeeverLubasch2023,KeeverLubasch2024} has demonstrated that classically optimized Hamiltonian simulation and compressed quantum circuits can outperform standard Trotter-Suzuki decompositions by several orders of magnitude. These approaches offer promising alternatives for optimizing the quantum circuits used in our simulation of lattice gauge theories.

\subsection{Preparing and Measuring Packaged Quantum States}
\label{sec:PrepAndMeasurement}

A central advantage of packaging (Sec.~\ref{sec:QLMTruncation}) is that excitations or superpositions thereof remain manifestly gauge invariant, simplifying both state preparation and error mitigation.

\begin{enumerate}
	\item \textbf{State Preparation:}
	\begin{itemize}
		\item \emph{Vacuum Initialization:} Begin in the gauge-invariant vacuum (the ground state with no excitations).  
		\item \emph{Create Excitations:} Apply gauge-invariant operators (e.g., $\hat{\psi}^\dagger \hat{U}\,\hat{\psi}$) to generate single- or multi-particle states.  
		\item \emph{Construct Superpositions:} Build hadronic wavefunctions or flux loops that remain within the physical subspace.
	\end{itemize}
	
	\item \textbf{Measurements:}
	\begin{itemize}
		\item \emph{Wilson Loops}: Observables like $\mathrm{Tr}(\prod \hat{U}_\ell)$ diagnose confinement.  
		\item \emph{Flux and Charge Distributions}: Probe how excitations are packaged into color-neutral bound states.  
		\item \emph{Gauge Constraint Checks}: Measuring $\hat{G}_x$ confirms the state remains in the physical subspace.
	\end{itemize}
\end{enumerate}

Because gauge-violating processes create unpackaged (unphysical) states, violations can be readily detected by checking whether $\hat{G}_x$ still leaves the state invariant.

\paragraph{Physical Subspace Projector.}
To formalize these checks, let $\hat{G}_x$ denote the local gauge transformation at site $x$.  Physical states $\ket{\Psi_{\mathrm{phys}}}$ satisfy
\[
\hat{G}_x\,\ket{\Psi_{\mathrm{phys}}} \;=\; \ket{\Psi_{\mathrm{phys}}}
\quad\forall\,x.
\]
Define the projector onto the physical subspace as
\[
\mathcal{P} \;=\; \prod_{x}\,\frac{1}{|G|}\,\sum_{g_x\in G}\,\hat{U}_{g_x},
\]
where $\hat{U}_{g_x}$ is the unitary representation of the gauge group element $g_x$ at site $x$, and $|G|$ is the normalized measure (or the group’s cardinality for a finite group).  Applying $\mathcal{P}$ to any state $\ket{\Psi}$ yields a gauge-invariant state $\ket{\Psi_{\mathrm{phys}}} = \mathcal{P}\ket{\Psi}$.

\paragraph{Error Operators and Gauge Violation.}
Now consider a generic error operator $\hat{E}$. If $[\hat{E},\hat{G}_x]\neq 0$ for some $x$, then acting on a physical state can produce a component outside the physical subspace:
\[
\mathcal{P}\,\hat{E}\,\ket{\Psi_{\mathrm{phys}}}
\;\neq\;
\hat{E}\,\ket{\Psi_{\mathrm{phys}}}.
\]
Hence, by periodically measuring $\hat{G}_x$ or employing circuits that project onto $\mathcal{P}$, one can detect (and mitigate) such gauge-violating noise.  Moreover, if \emph{all} excitations are packaged in gauge irreps, then operators attempting to break gauge symmetry become easier to detect because they drive the system into unphysical ``fractional'' states or ``partially'' entangled states.

\paragraph{Constructing Gauge-Invariant Logical Qubits.}
One can even encode logical qubits directly in $\mathcal{H}_{\mathrm{phys}}$.  For instance, consider two orthogonal physical states $\ket{\Psi_0}$ and $\ket{\Psi_1}$ satisfying
$\hat{G}_x\,\ket{\Psi_i}=\ket{\Psi_i}$.  
A logical qubit can be the span of these two basis states:
\[
\mathcal{C} \;=\; \mathrm{span}\{\ket{\Psi_0}, \ket{\Psi_1}\}
\;\subset\;
\mathcal{H}_{\mathrm{phys}}.
\]
Any error that would change the gauge quantum numbers (and thus remove the state from $\mathcal{C}$) is detectable by measuring $\hat{G}_x$.  In this sense, packaging offers an intrinsic ``gauge protection'' against certain errors, complementing standard quantum error correction \cite{Bombin2015} for noise that acts \emph{within} the physical subspace.

\subsection{Error Channels and Gauge Invariance}
\label{sec:ErrorChannelsGaugeConstraints}

Although packaging provides inherent robustness against gauge-violating processes, real devices experience a variety of noise and errors:

\begin{enumerate}
	\item \textbf{Gauge-Violating Errors:}  
	These errors move the state out of $\mathcal{H}_{\mathrm{phys}}$ by breaking local color or charge neutrality.  They can be suppressed or detected via:
	\begin{itemize}
		\item \emph{Penalty Terms:} Add $\lambda\sum_{x}(\hat{G}_x-I)^2$ to the Hamiltonian with large $\lambda$.  Violations acquire a high energy cost.
		\item \emph{Projective Checks:} Measure $\hat{G}_x$ (or a stabilizer encoding it) at intervals to detect unphysical states.
	\end{itemize}
	
	\item \textbf{Gauge-Allowed Noise:}  
	Errors that preserve $\hat{G}_x$ (e.g., certain phase flips or dephasing) do \emph{not} break Gauss’s law but still corrupt the state within $\mathcal{H}_{\mathrm{phys}}$.  Standard quantum error correction or error-mitigation techniques are required to handle these.
	
	\item \textbf{Gauge-Invariant Codes:}  
	Encoding logical qubits in color-singlet subspaces ($\mathcal{H}_{\mathrm{phys}}$) automatically flags gauge-violating errors as ``leakage''.  While gauge invariance alone does not fix within-sector noise, it provides a first layer of protection and a simpler route to detecting unphysical processes.
\end{enumerate}

In summary, although ``packaging'' cannot eliminate all errors, it naturally restricts the types of noise that must be actively corrected and simplifies the detection of gauge-violating events.

Subsequent sections will show explicit examples for particular gauge groups (e.g., $\mathrm{U}(1)$, $\mathrm{SU}(2)$, $\mathrm{SU}(3)$) and demonstrate how multi-particle packaged states serve as building blocks for exploring non-perturbative physics on quantum simulators.

\section{Applications of Packaged Quantum States: A General Perspective}
\label{sec:ApplicationsOfPQS0}

In previous sections, we introduced packaged quantum states - states in which each single excitation (flux quantum, quark, etc.) is carried by a full irrep of the local gauge group.  Such packaging ensures that no partial or fractional internal quantum numbers (IQNs) can appear, in accordance with local gauge invariance.  

Thus far, we have mostly focused on single-particle packaging and non-entangled multi-particle states, which - while crucial for obeying Gauss’s law - does not itself produce internal entanglement. However, as soon as multiple excitations (irreps) combine within one gauge-invariant superselection sector, we can realize \textbf{nontrivial packaged entanglement} between them. These multi-particle packaged entangled states are often the key to understanding color-singlet hadrons, flux-loop resonances, and other strongly correlated physics in lattice gauge theories (LGTs).  

In this section, we (1) review how packaging arises from gauge invariance and superselection, (2) discuss how multi-particle states can become nontrivially packaged entangled while still obeying all local constraints, and (3) explore how such packaged entangled states are prepared, probed, and potentially leveraged for error mitigation.  We do so without committing to a specific gauge group ($\mathrm{U}(1)$, $\mathrm{SU}(2)$, $\mathrm{SU}(3)$, etc.), as the principles hold broadly.

\subsection{Gauge Invariance and Packaging Principle}

Gauge invariance (or local symmetry) is the requirement that a theory remain invariant under local transformations that can vary from point to point in spacetime.  Crucially, this means each physical particle or flux excitation must transform as a full irrep of the local gauge group $G$.  In an Abelian model such as $\mathrm{U}(1)$, a single excitation carries one unit of charge; in a non-Abelian model (e.g., $\mathrm{SU}(2)$, $\mathrm{SU}(3)$), each excitation carries a fundamental or higher representation.  

On a deeper level, insisting on local gauge invariance \emph{forces} the existence of a gauge field (and hence an interaction). This is the essence of the gauge principle: local gauge symmetry is not merely a mathematical device; it is a guiding principle that dictates how interactions must be incorporated into the theory.

An important consequence of local gauge invariance is what we call the \textbf{packaging principle}: the full set of IQNs - such as color, electric charge, etc. - of each particle is packaged into a single-particle operator and cannot be fractionally distributed.  In other words, one cannot split IQNs into partial (sub-irrep) excitations.

To see this at the operator level, consider the matter-field operator $\hat{\psi}^\dagger_{x,\alpha}$ at lattice site (or spacetime point) $x$.  Its gauge index $\alpha$ runs over the components of some representation $\mathbf{r}$ of $G$.  Under a local gauge transformation $g_x\in G$, the operator transforms as
\[
\hat{\psi}^\dagger_{x,\alpha} \;\;\mapsto\;\; \sum_{\beta}\,D_{\alpha\beta}(g_x)\,\hat{\psi}^\dagger_{x,\beta},
\]
where $D_{\alpha\beta}(g_x)$ is the representation matrix.  This demonstrates that $\hat{\psi}^\dagger_{x,\alpha}$ cannot be factorized into smaller color- or charge-carrying pieces.  Once such an excitation is created, \emph{all} of its IQNs remain inseparably bound together, forming an irreducible block.

Hence, the packaging principle can be summarized in two key points:
\begin{enumerate}
	\item The IQNs cannot be split into fractions; the creation operator itself is an irrep.
	
	\item Once a particle is created, all of its IQNs must be treated as a single ``package''. In packaged entangled states, all IQNs are entangled together and cannot be separated.
\end{enumerate}

This indivisible nature of excitations under the gauge group is both a manifestation of local gauge invariance and a fundamental hallmark of the theory’s internal structure.

\subsection{Packaged Non-entangled States}

A single-particle state at site $x$ is created by 
\begin{equation}\label{SingleParticleState}
	\ket{\psi} 
	\;=\; 
	\hat{\psi}^\dagger_{x,\alpha}\,\ket{0}.
\end{equation}
Here, $\hat{\psi}^\dagger_{x,\alpha}$ spans a full irrep of the gauge group, carrying its entire IQNs (e.g., color or charge). Because there is only one excitation, no multi-particle entanglement arises. The IQNs cannot be split among partial excitations; they are packaged in this single operator.

Multiple excitations can be created at different sites, each by a single creation operator carrying a full gauge irrep.  A general (non-entangled) multi-particle state can be written as a product of single-particle states, for instance,
\[
\ket{\Psi}
\;=\;
\Bigl(
\hat{\psi}^\dagger_{x_1,\alpha_1}
\hat{\psi}^\dagger_{x_2,\alpha_2}
\cdots
\hat{\psi}^\dagger_{x_k,\alpha_k}
\Bigr)
\ket{0},
\]
where each $\hat{\psi}^\dagger_{x_i,\alpha_i}$ creates one packaged excitation.  Although one may form more general superpositions, in the \emph{non-entangled} case each particle remains a separate, irreducible package.  Crucially, gauge invariance requires the total state to lie in an appropriate overall gauge representation (often the singlet sector), ensuring that quantum numbers are packaged correctly while remaining unentangled.

\subsection{Superposition and Superselection Rules}

In gauge theories, certain global quantum numbers cannot be altered by any \emph{local}, gauge-invariant operator.
Two main examples are: \textbf{total (net) gauge charge} (e.g., electric charge, color charge) and \textbf{global (topological) flux or winding} (e.g., integer flux threading a torus).

By Gauss’s law, the net gauge charge in any physical state is fixed.  Similarly, on a lattice with periodic boundary conditions, the global flux threading each non-contractible loop is also integer-valued and typically cannot be changed by local operations.  Consequently, the Hilbert space is divided into disjoint superselection sectors labeled by these conserved quantum numbers.  No gauge-invariant local operator can connect different sectors, which forbids coherent superpositions of, for example, ``$\text{net charge} = 0$'' and ``$\text{net charge} = 1$.''  

Nonetheless, \emph{within} a single sector, one can form nontrivial superpositions and even entangled states (often called \emph{packaged entangled states}), as long as they respect the global gauge constraints.  We now discuss each type of superselection in more detail.

\subsubsection{Charge Superselection}

\paragraph{Gauss’s Law and Fixed Total Charge.}
Gauge theories impose that the total gauge charge is fixed in each physical sector \cite{WWW1952,DHR1,DHR2,StreaterWightman}. By Gauss’s law, one cannot locally insert or remove a net charge; operators that create gauge excitations must do so in charge-conjugate pairs or in a way that net charge remains unchanged.  This yields a superselection rule: no physical process allows transitions between distinct total-charge sectors.

\paragraph{No Cross-Sector Superpositions.}
Since no local gauge-invariant operator can connect states of different net charge, any formal superposition such as
\[
\alpha\,\ket{Q=0} \;+\; \beta\,\ket{Q=1}
\]
is \emph{unphysical} in the sense that no measurement or local process can reveal a relative phase between $\ket{Q=0}$ and $\ket{Q=1}$.  Operationally, it behaves like a classical mixture, so the charge is effectively a ``good'' quantum number labeling disjoint superselection sectors.

\paragraph{Within-Sector Superpositions.}
Although you cannot superpose different net charges, it is possible to form superpositions within a single charge sector - leading to interesting multiparticle or flux-loop configurations.  In that sense, ``packaging'' means that each particle or gauge excitation carries its full irrep (IQNs cannot be split), yet multiple such excitations can still become packaged entangled as long as the overall net charge remains fixed.

\subsubsection{Winding (Topological) Superselection}

\paragraph{Global Flux in Pure Gauge Theories.}
In many lattice gauge theories with periodic boundary conditions (e.g., on a torus), one can classify states by the integer flux (``winding number'') threading each non-contractible loop \cite{Polyakov1975,Luscher1980}.
Denote these windings by
\[
W_x \;=\; \sum_{\ell \in \text{loop}_x} E_\ell, 
\quad 
W_y \;=\; \sum_{\ell \in \text{loop}_y} E_\ell,
\]
and so on for higher dimensions.  Here, $E_\ell$ is an electric flux quantum number on link $\ell$.  If there are no external charges and the boundary is fully periodic, these winding numbers are typically \emph{globally} conserved by local dynamics.

\paragraph{Why Winding is Superselected.}
A local operator acting on a finite region cannot ``cut'' or ``re-thread'' an entire flux line around a non-contractible loop, so it cannot change $W_x$ or $W_y$.  Gauss’s law also requires flux lines either to form closed loops or to terminate on charges; hence, in a purely periodic system with no charges, different winding sectors $\ket{W=0}, \ket{W=1}, \dots$ cannot mix via local gauge-invariant processes.  From a practical standpoint, this is analogous to charge superselection: one cannot have a coherent superposition $\alpha\,\ket{W=0} + \beta\,\ket{W=1}$ that displays interference under local measurements.

\paragraph{Instantons and Large Gauge Transformations.}
In certain models, especially in lower dimensions or with special boundary conditions, nonlocal processes (sometimes called instantons or large gauge transformations) can connect different windings \cite{Belavin1975}. If these processes are non-negligible, winding may cease to be an \emph{exact} superselection quantum number.  Nonetheless, in many physically relevant settings - especially at large volume or in higher-dimensional lattice gauge theories - tunneling amplitudes between winding sectors are negligible, leaving winding effectively superselected.

\paragraph{Flux-Loop States.}
In \emph{pure} gauge theories (with no matter fields), excitations manifest as closed flux loops or topologically nontrivial flux configurations.  A schematic flux-loop state might look like
\[
\ket{\Psi_{\text{flux}}}
\;=\;
\sum_{\{r_\ell\}} 
\beta(\{r_\ell\})
\bigotimes_{\ell}
\ket{r_\ell},
\]
where $\ket{r_\ell}$ are basis states on each link $\ell$, and $\beta(\{r_\ell\})\neq 0$ only if each vertex satisfies Gauss’s law.  Such states reside in a particular winding sector (or sometimes a superposition of flux loops \emph{within} one sector) and illustrate how ``packaging'' enforces that the flux must close on itself or connect to charges in a gauge-invariant way.

\subsection{Packaged Entangled States}

\paragraph{Definition and Motivation.}
Packaged entangled states \cite{Ma2025} arise when multiple irreps (single-particle excitations) are \emph{superposed} in a single gauge sector, all subject to Gauss’s law.  While each single-particle operator is packaged (carrying a full irrep of the gauge group and thus inseparable IQNs), entanglement emerges from how these irreps combine or superpose across different sites or degrees of freedom.  This multi-particle packaged entanglement underlies much of the rich physics in gauge theories, from color-singlet hadrons in QCD to flux-tube interference in pure gauge theories.

\paragraph{From Single-Particle to Multi-Particle States.}
A single packaged excitation,
\[
\ket{\Psi_{\rm single}} \;=\; \hat{\psi}^\dagger_{x,\alpha}\,\ket{0},
\]
creates one irreps index $\alpha$ at site $x$.  No multi-particle entanglement arises because there is only one excitation.  In contrast, a multi-particle state
\[
\ket{\Psi} 
\;=\; 
\sum\nolimits_n \,\alpha_n\, 
\bigl(\hat{\psi}^\dagger_{x_1,\alpha_1}\cdots \hat{\psi}^\dagger_{x_k,\alpha_k}\bigr)\ket{0}
\]
can exhibit nontrivial superpositions provided all excitations together form a gauge-invariant configuration (e.g.\ net color singlet or zero net charge).  Within that single sector, one may have nontrivial packaged entanglement among the excitations’ indices or spatial locations.

\paragraph{Composite Gauge-Invariant States.}
To ensure gauge invariance, each physical state must remain invariant under local transformations.  Let $\hat{\psi}^\dagger_{x,i}$ create an irrep index $i$ at site $x$.  A multi-particle wavefunction then reads
\begin{equation}\label{eq:general_state_app}
	\ket{\Psi} 
	= 
	\sum_{\{i_k\}} \alpha_{i_1 i_2 \cdots i_N}\, 
	\hat{\psi}^\dagger_{x_1,i_1}
	\hat{\psi}^\dagger_{x_2,i_2}
	\cdots 
	\hat{\psi}^\dagger_{x_N,i_N}\,\ket{0},
\end{equation}
where the coefficients $\alpha_{i_1 i_2 \cdots i_N}$ must form an \emph{invariant tensor} under the gauge group $G$.  In physical terms, a color-singlet meson or a flux-loop state is exactly such a gauge-invariant combination of packaged excitations.

\begin{example}[Mesons/Baryons in Non-Abelian Gauge Theories]
	In $\mathrm{SU}(3)$ QCD, quarks carry the fundamental representation $\mathbf{3}$ and antiquarks carry $\overline{\mathbf{3}}$.  A color-singlet meson can be written as
	\[
	\sum_{\alpha=1}^3 
	\hat{q}^\dagger_{x,\alpha}\,\hat{\bar{q}}^\dagger_{y,\alpha}\,\ket{0},
	\]
	where the color indices $\alpha$ of the quark and antiquark are packaged entangled to ensure overall gauge invariance. Baryons, composed of three quarks in an antisymmetric color combination, form more elaborate packaged entangled states.  Crucially, no local gauge-invariant operator can shift the total color charge from 0 to 1, so color singlet is a superselected sector. Within that sector, rich multi-particle packaged entanglement emerges.
\end{example}

\begin{example}[Flux Loops in Pure Gauge Theories]
	In a pure $\mathrm{U}(1)$ lattice gauge theory with periodic boundary conditions, physical excitations can manifest as flux lines or loops.  States are often labeled by a winding number $(W_x,W_y,\dots)$. Even if different windings are superselected globally, one can still form superpositions of flux-loop configurations within a single winding sector:
	\[
	\ket{\Psi_{\text{flux}}} 
	= 
	\sum_{\{e_\ell\} \in \text{loop}} 
	\beta(\{e_\ell\}) 
	\bigotimes_{\ell} 
	\ket{e_\ell},
	\]
	where the link-flux states $\ket{e_\ell}$ form closed loops or net winding, satisfying Gauss’s law at each site. The nontrivial packaged entanglement appears in the coefficients $\beta(\{e_\ell\})$, allowing interference among different loop shapes or positions.
\end{example}

\paragraph{Why Packaged Entanglement Matters.}
While each single-particle creation operator is packaged (carrying a complete irrep by itself), multi-particle combinations enable packaged entanglement across different irreps or spatial sites. This packaged entanglement is at the heart of phenomena like hadron formation, flux-tube interference, and potential quantum simulation or quantum information applications of lattice gauge theories.

\paragraph{Caveats: $\theta$-Vacua and Topological Superselection.}
In some continuum field theories (e.g.\ QCD with a $\theta$-term), one can form ``$\theta$-vacua'' as superpositions of different topological winding sectors.  However, these often require nonlocal processes or instanton tunneling \cite{Belavin1975}.  In a finite lattice or large-volume limit with local dynamics, tunneling amplitudes between distinct windings may be negligible, effectively superselecting each winding sector.  Thus, while global flux or topological charge can sometimes mix via boundary or instanton effects in continuum settings, on a discrete lattice with strictly local gauge-invariant operators, such mixing is often forbidden or heavily suppressed.

\subsection{Preparing and Probing Multi-Particle Packaged Entangled States}
\label{sec:PreparationProbing}

Multi-particle packaged entangled states lie at the heart of many interesting phenomena in lattice gauge theories, from color-singlet hadrons (mesons, baryons) to flux-tube superpositions.  Once such states are prepared, one also needs practical methods to measure their properties and to maintain gauge invariance against errors.  Below we outline common protocols for creating these states, the observables used to probe them, and strategies for error mitigation and circuit design.

\subsubsection{State Preparation Protocols}

Several standard strategies exist for preparing multi-particle packaged entangled states in a gauge-invariant framework:

\paragraph{(1) Gate-by-Gate Construction (Digital Simulation).}
\begin{itemize}
	\item Start with a gauge-invariant vacuum (e.g., no matter and zero flux).
	\item Apply local \emph{gauge-covariant} gates that create or move excitations while respecting Gauss’s law.  For example, create a quark-antiquark pair on neighboring sites, then separate them with hopping operations.  
	\item The resulting state can be a color-singlet meson or a superposition of flux loops, automatically remaining within the physical (gauge-invariant) subspace.
\end{itemize}

\paragraph{(2) Adiabatic Preparation (Analog or Digital).}
\begin{itemize}
	\item Initialize the system in a simple limit (e.g., strong coupling), where flux tubes \cite{Bali1995,Bali2001,Takahashi2002,Zohar2011} or quark pairs are localized and easy to prepare.
	\item Slowly tune the coupling or other parameters to the desired regime (e.g., weaker coupling).
	\item If done adiabatically, the state remains in a low-energy, gauge-invariant subspace and develops nontrivial packaged entanglement characteristic of the target Hamiltonian.
\end{itemize}

\paragraph{(3) Global Quench (Analog Platforms).}
\begin{itemize}
	\item Prepare a simple product state (e.g., all links in $\ket{0}$, no flux).
	\item Suddenly switch on the full gauge-invariant Hamiltonian.
	\item Under time evolution, the system can dynamically generate superpositions of multi-particle excitations, such as emergent flux tubes or hadrons.
\end{itemize}

\subsubsection{Probing Packaged Entanglement and Observables}

Once a multi-particle packaged entangled state is prepared, one can measure various observables to characterize the physics and quantify entanglement:

\paragraph{Wilson Loops.}
Evaluate $W(\mathcal{C}) = \mathrm{Tr}\bigl(\prod_{\ell\in \mathcal{C}} U_\ell\bigr)$ along closed contours $\mathcal{C}$.  These help detect confinement (area vs.\ perimeter law) and reveal flux-tube formation.

\paragraph{Local Flux Operators.}
Measure link electric fields $E_\ell$ to see how flux lines or loops are distributed.  In pure gauge theories, such measurements clarify how flux loops form and evolve.

\paragraph{Hadronic Correlators.}
Construct color-singlet operators such as $\hat{q}^\dagger \hat{\bar{q}}^\dagger$ (mesons) or three-quark combinations (baryons) to measure correlation functions and extract mass spectra.

\paragraph{Packaged Entanglement Entropy.}
For smaller systems, one can perform (full or partial) state tomography to calculate reduced density matrices and quantify packaged entanglement entropy. Alternatively, entanglement witnesses can bound the entanglement in larger systems without full tomography.

\subsubsection{Error Mitigation and Gauge-Invariant Codes}

A key advantage of the packaged approach is its built-in resilience to gauge-violating errors.  Physical states must satisfy
\[
\hat{G}_x \ket{\Psi_{\mathrm{phys}}} = \ket{\Psi_{\mathrm{phys}}}\quad 
\forall\,x,
\]
where $\hat{G}_x$ is the generator of local gauge transformations at site $x$.  One can define the projector onto this physical subspace as 
\[
\mathcal{P} = \prod_{x}\,\frac{1}{|G|}\sum_{g_x \in G}\,\hat{U}_{g_x},
\]
which enforces gauge invariance.  Any local error operator $\hat{E}$ that fails to commute with $\hat{G}_x$ drives the state out of $\mathcal{H}_{\mathrm{phys}}$, making it detectable as a leakage error.

In practice, \emph{digital} quantum simulations often incorporate:
\begin{itemize}
	\item \textbf{Syndrome Measurements:} Periodic checks of $\hat{G}_x$ to ensure the system has not left the gauge-invariant subspace.
	\item \textbf{Gauge-Invariant Circuits:} Designing gates $U$ that commute with local gauge transformations, so the evolution stays within $\mathcal{H}_{\mathrm{phys}}$.
	\item \textbf{Logical Encoding:} Logical qubits encoded as superpositions of gauge-invariant states, $\ket{\Psi_L} = c_0 \ket{\Psi_0} + c_1 \ket{\Psi_1}$, naturally flag or suppress errors that attempt to break gauge invariance.
\end{itemize}
Hence, multi-particle packaged states can form the basis of gauge-invariant quantum error-correcting codes, where standard QEC methods handle gauge-preserving errors, and ``unphysical'' gauge-violating errors are readily detected.

\subsubsection{Digital Circuit Implementations}

To realize multi-particle packaged entangled states on a digital quantum platform, one constructs circuits whose local gates are gauge-covariant. Formally, a gate $U$ acting on site (or link) $x$ must satisfy
\[
U\,D(g_x) \;=\; D(g_x)\,U\quad \forall\,g_x \in G,
\]
where $D(g_x)$ is the representation of $g_x$.  This ensures that if the system starts in a gauge-invariant state, it remains in that subspace after applying $U$.  For instance, constructing a color-singlet meson state might involve:
\begin{itemize}
	\item Initializing the vacuum in $\mathcal{H}_{\mathrm{phys}}$.
	\item Applying local creation gates that produce a quark-antiquark pair, each carrying an irreps index, then ``contracting'' them via an invariant tensor (such as a Kronecker delta or antisymmetric symbol) to form a singlet.
\end{itemize}

By carefully designing these operations, one obtains robust, gauge-invariant packaged entangled states (e.g., mesons, baryons, flux-loop superpositions) that remain protected from certain gauge-violating errors.

\subsection{Advantages of Packaged Entangled States}

Multi-particle \emph{packaged entangled states} are central to nonperturbative physics in lattice gauge theories (LGTs).  While single-particle packaging ensures each excitation carries a full irrep of the gauge group, it is multi-particle packaging and entanglement that capture phenomena such as color confinement \cite{Gross1973,Politzer1973}, flux-tube formation, and hadron structure.  Below, we highlight the physical, computational, and error-mitigation advantages of these states.

\paragraph{1. Nontrivial Correlations and Gauge-Invariant Packaged Entanglement.}
\begin{itemize}
	\item \textbf{Collective Phenomena:}
	Many hallmark features of gauge theories - \emph{confinement, hadron spectra, flux-tube dynamics} - arise from collective correlations among multiple excitations.  Single-particle irreps alone cannot capture color binding or multi-particle flux loops.
	
	\item \textbf{Meaningful Packaged Entanglement Measures:}
	Packaged entanglement in a gauge theory must respect Gauss’s law and superselection rules. Multi-particle packaged states provide a natural, gauge-invariant basis for studying how quark-antiquark (or flux-loop) packaged entanglement and correlations spread across the lattice.
	
	\item \textbf{Real-Time Dynamics:}
	Meson scattering, baryon formation, and other multi-particle processes inherently involve entangled excitations interacting via gauge fields.  Quantum simulators - especially those based on packaged states - are uniquely suited to tackle real-time evolution in regimes where classical methods (e.g., Monte Carlo) encounter sign problems.
\end{itemize}

\paragraph{2. Automatic Gauss’s Law Enforcement and Error Suppression.}
\begin{itemize}
	\item \textbf{Built-In Gauge Invariance:}
	Because each excitation is a full gauge-irrep (e.g., $\mathbf{3}$ vs.\ $\overline{\mathbf{3}}$ in $\mathrm{SU}(3)$), forming a color singlet or net-zero charge configuration is straightforward.  One avoids ``accidental'' gauge violations that might otherwise arise when partially filling irreps.
	\item \textbf{Superselection Protection:}
	Cross-sector interference (e.g., net charge $0$ vs.\ net charge $1$) is forbidden by gauge theory.  Packaged states remain safely in their correct sector; errors that would create a fractional charge or partial flux become gauge-violating and are thus readily detectable.
	\item \textbf{Error Detectability:}
	If any local operator tries to shift the system out of the gauge-invariant subspace (breaking Gauss’s law), that error can be exposed as a leakage into unphysical states. This synergy of gauge invariance and standard quantum error correction (QEC) \cite{Bombin2015} bolsters robustness for quantum simulations of LGTs.
\end{itemize}

\paragraph{3. Capturing Hadronic and Flux-Loop Excitations.}
\begin{itemize}
	\item \textbf{Hadrons in non-Abelian Theories:}
	In $\mathrm{SU}(3)$ lattice QCD \cite{DeGrandDeTar2006}, quarks (in $\mathbf{3}$) and antiquarks (in $\overline{\mathbf{3}}$) must combine into color singlets.  Mesons (quark-antiquark) or baryons (three quarks) emerge naturally as entangled superpositions of packaged excitations.  
	\item \textbf{Flux Loops in Abelian Theories:}
	Pure $\mathrm{U}(1)$ gauge theories constrain flux lines to close on themselves or connect charges.  Packaged excitations ensure integer flux quanta, allowing superpositions of flux loops or winding numbers that respect local constraints at every site.
\end{itemize}

\paragraph{4. Single-Particle vs.\ Multi-Particle Packaging.}
\begin{itemize}
	\item \textbf{Trivial Single-Particle Entanglement:}
	A single packaged excitation (one irreps index) does not exhibit packaged entanglement by itself. It is, however, still irreducible with respect to the gauge group.
	
	\item \textbf{Rich Multi-Particle Packaged Entanglement:}
	When two or more irreps combine in a gauge-invariant sector, they can form \emph{packaged entangled states} (e.g., meson-like quark-antiquark superpositions).  These states are the essential resource for capturing nonperturbative physics such as confinement or topological flux loops.
\end{itemize}

\paragraph{5. Practical Impact and Outlook.}
\begin{itemize}
	\item \textbf{Cleaner Simulations and Reduced Overhead:}
	By eliminating fractional charges or unphysical excitations \emph{from the outset}, one reduces the size of the relevant Hilbert space and avoids large penalty terms that would otherwise enforce gauge constraints.
	\item \textbf{Implementation via Quantum Link Models:}
	Finite spin (or qudit) truncations on each link naturally accommodate packaged excitations.  Truncated representations still respect local gauge invariance, neatly aligning with digital and analog quantum hardware.
	
	\item \textbf{Potential Applications:}
	\begin{enumerate}
		\item \emph{Lattice QCD Simulations:} \cite{DeGrandDeTar2006} Study hadron spectroscopy, confinement, and real-time scattering in $\mathrm{SU}(3)$ or $\mathrm{SU}(2)$ gauge theories.
		\item \emph{Confinement-Deconfinement Transitions:} Measure Wilson loops or flux observables to identify phase transitions in $\mathrm{SU}(N)$ or $\mathrm{U}(1)$ models.
		\item \emph{Sign-Problem-Free Real-Time Evolution:} Track multi-particle packaged states dynamically in regimes inaccessible to classical Monte Carlo.
	\end{enumerate}
	
	\item \textbf{Gauge Codes and Fault Tolerance:}\cite{Bombin2015}
	Multi-particle packaged states allow one to encode logical qubits in the gauge-invariant subspace, improving detectability of gauge-violating errors.  Combined with standard QEC, this ``gauge code'' approach aids fault-tolerant quantum simulations of LGTs.
\end{itemize}

In the subsequent sections, we will illustrate these concepts for specific gauge groups ($\mathrm{U}(1)$, $\mathrm{SU}(2)$, $\mathrm{SU}(3)$), showing how no partial flux or color emerges and how multi-particle packaged entangled states embody the core of hadronic and flux-loop physics in lattice gauge theories.

\section{$U(1)$ Lattice Gauge Theory}
\label{sec:U1Gauge}

We now focus on the simplest gauge group, $\mathrm{U}(1)$.\cite{Feynman1949}  Despite its apparent simplicity compared to non-Abelian groups, a $\mathrm{U}(1)$ lattice gauge theory \cite{Wilson1974,Polyakov1975,Wegner1971} already demonstrates the key packaging principle: all excitations are integer flux quanta forming closed loops.  In 2D with periodic boundaries, these flux lines can wind around the torus, leading to topological superselection sectors.  This model also serves as a foundational testbed for quantum simulators (e.g.\ compact QED, or the Schwinger model in lower dimensions), illustrating how gauge invariance enforces integer flux packaging.

\subsection{2D Lattice Setup for $\mathrm{U}(1)$ Gauge}

We focus on a two-dimensional square lattice of size $L \times L$ with periodic boundary conditions, so it can be viewed topologically as a torus.  Each lattice site is labeled by integer coordinates
\[
\mathbf{n} = (n_x, n_y), 
\quad
n_x, n_y \in \{0,1,\dots,L-1\}.
\]
At each site $\mathbf{n}$, there are two outgoing links:
\begin{enumerate}
	\item A link in the $\hat{x}$ direction, denoted $\ell_{(\mathbf{n}, \hat{x})}$,
	\item A link in the $\hat{y}$ direction, denoted $\ell_{(\mathbf{n}, \hat{y})}$.
\end{enumerate}
With periodic boundaries, any link exiting the lattice on one edge reenters on the opposite edge.  Consequently, the total number of links is $2L^2$.  

In a $\mathrm{U}(1)$ gauge theory, excitations carry integer flux.  Although $\mathrm{U}(1)$ lacks the non-Abelian ``color'' structure, the principle of \emph{packaging} still applies: flux must come in integer units, so partial or fractional flux excitations cannot occur.  Physically, this setup is often used because:
\begin{itemize}
	\item \textbf{Topological Winding:} Periodic boundaries allow global flux to wrap around the lattice in the $x$ and/or $y$ directions, illustrating superselection by winding number.
	\item \textbf{Simplicity in 2D:} Tracking how flux lines close into loops is intuitive, yet the same core ideas (Gauss’s law, link variables, integer flux) extend naturally to higher dimensions.
\end{itemize}

Hence, even in the simplest $\mathrm{U}(1)$ case, the packaging principle enforces that flux excitations appear in whole-integer increments and form closed loops or connect to charged matter fields, reflecting local gauge invariance at every lattice site.

\subsection{Hamiltonian of Pure $\mathrm{U}(1)$}

In the Kogut-Susskind formulation of a pure $\mathrm{U}(1)$ gauge theory (in 2+1D or 3+1D), the Hamiltonian takes the form (see Eq.(\ref{GaugeTheoryHamiltonians}))
\begin{equation}\label{U1KogutSusskindHamiltonian_app}
	\hat{H}
	\;=\;
	\underbrace{\frac{g^2}{2}\,\sum_{\ell}\hat{E}_\ell^2}_{\text{electric term}}
	\;+\;
	\underbrace{\frac{1}{2g^2}\,\sum_{\Box}\Bigl(\hat{U}_{\Box} + \hat{U}_{\Box}^\dagger\Bigr)}_{\text{magnetic (plaquette) term}},
\end{equation}
where:
\begin{itemize}
	\item $\ell$ runs over all links,
	\item $\hat{E}_\ell$ is the integer-valued electric field operator on link $\ell$,
	\item $\hat{U}_\Box = \hat{U}_{\ell_1}\hat{U}_{\ell_2}\hat{U}_{\ell_3}^\dagger\hat{U}_{\ell_4}^\dagger$ 
	is the product of link operators ($\hat{U}_\ell = e^{i \hat{\theta}_\ell}$) around a plaquette $\Box$.
\end{itemize}
This Hamiltonian commutes with the gauge transformation generators $\hat{G}_{\mathbf{n}}$ at every site $\mathbf{n}$, ensuring that time evolution remains within the physical subspace satisfying Gauss’s law.

\paragraph{Electric Term.}
\[
\frac{g^2}{2}\sum_{\ell}\hat{E}_\ell^2
\]
penalizes large flux $\lvert e_\ell\rvert$.  At strong coupling ($g^2 \to \infty$), the vacuum is dominated by low-flux (often zero-flux) configurations.  Physically, this term reflects the energy stored in the electric field.

\paragraph{Magnetic (Plaquette) Term.}
\[
-\;\frac{1}{2g^2}\sum_{\Box}\bigl(\hat{U}_\Box + \hat{U}_\Box^\dagger\bigr)
\]
encourages configurations where the product of link phases around each plaquette is close to 1 (i.e., minimal magnetic flux).  It can create quantum superpositions of flux-loop states by allowing local plaquette ``flips'' that add or remove small loops of flux.  At weaker coupling ($g^2 \ll 1$), these fluctuations can delocalize flux lines, generating resonant superpositions of multiple loop configurations.

\paragraph{Dynamics and Confinement.}
By tuning $g^2$, one can explore different regimes:
\begin{itemize}
	\item \emph{Strong coupling} ($g^2$ large): Flux configurations are localized; the electric term dominates.
	\item \emph{Weak coupling} ($g^2$ small): The magnetic term enables large quantum fluctuations, potentially leading to deconfinement or more packaged entangled flux-loop states (depending on dimensionality and boundary conditions).
\end{itemize}
Because the Hamiltonian respects local gauge invariance, all physical excitations must be gauge-invariant, manifesting as integer flux loops (packaged excitations) or global windings in the $\mathrm{U}(1)$ case.

\subsection{Local $\mathrm{U}(1)$ Gauge Transformations and Gauss’s Law}
\label{sec:U1_GaussLaw}

In a \emph{pure} $\mathrm{U}(1)$ lattice gauge theory (i.e., without matter fields), local gauge transformations and Gauss’s law enforce that electric flux can only form closed loops or wrap around the periodic boundaries.  

\paragraph{Local Gauge Transformations.}
At each site $\mathbf{n} = (n_x,n_y)$, a gauge transformation is specified by a phase $\alpha_{\mathbf{n}} \in [0,2\pi)$.  This transformation shifts the link angles ($\hat{\theta}_\ell$) on outgoing and incoming links in opposite directions.  Concretely,
\[
\hat{\theta}_{(\mathbf{n}, \hat{\mu})} 
\;\longrightarrow\; 
\hat{\theta}_{(\mathbf{n}, \hat{\mu})} + \alpha_{\mathbf{n}},
\quad
\hat{\theta}_{(\mathbf{n}-\hat{\mu}, \hat{\mu})}
\;\longrightarrow\;
\hat{\theta}_{(\mathbf{n}-\hat{\mu}, \hat{\mu})} 
- \alpha_{\mathbf{n}},
\]
where $\hat{\mu}$ denotes a unit vector $\hat{x}$ or $\hat{y}$.  The electric field operators $\hat{E}_\ell$ transform correspondingly to preserve the canonical commutation relations.

\paragraph{Gauss’s Law Constraint.}
With no matter fields present, Gauss’s law at each site $\mathbf{n}$ demands zero net electric flux:
\[
\bigl(\hat{E}_{(\mathbf{n},\hat{x})}
+ \hat{E}_{(\mathbf{n},\hat{y})}\bigr)
\;-\;
\bigl(\hat{E}_{(\mathbf{n}-\hat{x},\hat{x})}
+ \hat{E}_{(\mathbf{n}-\hat{y},\hat{y})}\bigr)
= 0.
\]
Equivalently, any flux entering site $\mathbf{n}$ must exit it.  Denoting $\hat{G}_{\mathbf{n}}$ as the operator that implements the gauge transformation at $\mathbf{n}$, physical states $\ket{\Psi_{\mathrm{phys}}}$ satisfy
\begin{equation}\label{Eq:U1GaussLaw}
	\hat{G}_{\mathbf{n}}\,\ket{\Psi_{\mathrm{phys}}} 
	\;=\;
	\ket{\Psi_{\mathrm{phys}}}, 
	\quad \forall \mathbf{n}.
\end{equation}
In the flux eigenbasis $\ket{\{e_\ell\}}$, this requirement translates to
\[
\sum_{\ell \,\in\, \text{out}(\mathbf{n})} e_\ell
\;-\;
\sum_{\ell \,\in\, \text{in}(\mathbf{n})} e_\ell
= 0
\quad
(\forall\,\mathbf{n}),
\]
meaning only loop configurations (or global windings on the torus) are allowed.  No net source or sink of flux can appear.  As a result, the physical subspace $\mathcal{H}_{\mathrm{phys}}$ is spanned by states with closed flux lines or wrapped flux consistent with periodic boundary conditions.

\subsection{Physical (Gauge-Invariant) Hilbert Subspace}

The total (unconstrained) Hilbert space is
\[
\mathcal{H}
\;=\;
\bigotimes_{\ell=1}^{2L^2}
\mathcal{H}_\ell,
\]
where each $\mathcal{H}_\ell$ is spanned by the basis $\{\ket{e_\ell}\}$ with $e_\ell\in\mathbb{Z}$.  Thus,
\[
\hat{E}_\ell \,\ket{e_\ell}
\;=\;
e_\ell \,\ket{e_\ell}.
\]

A general state $\ket{\Phi}\in\mathcal{H}$ can be written as
\[
\ket{\Phi}
\;=\;
\sum_{\{e_\ell\}} 
C(\{e_\ell\})
\,\bigotimes_{\ell}\ket{e_\ell},
\]
where $\{e_\ell\}$ runs over all integer flux configurations on each link.

Physical Subspace $\mathcal{H}_{\text{phys}}$ is defined by Gauss’s law:
$
\hat{G}_{\mathbf{n}}\ket{\Psi_{\text{phys}}}
= \ket{\Psi_{\text{phys}}}, ~ \forall \mathbf{n},
$
i.e., $\hat{G}_{\mathbf{n}}=1$ at each site $\mathbf{n}$ means only those configurations with net zero flux at each site survive.  Formally,
\[
\mathcal{H}_{\text{phys}}
\;=\;
\Bigl\{
\ket{\Psi}\in \mathcal{H}
\;\Big\vert\;
\hat{G}_{\mathbf{n}}\ket{\Psi}=\ket{\Psi},
\;\forall \mathbf{n}
\Bigr\}.
\]

In the flux basis, a necessary condition for $\ket{\Psi}$ to be in $\mathcal{H}_{\text{phys}}$ is that any nonzero amplitude $C(\{e_\ell\})$ must have
\[
\sum_{\ell\in\text{out}(\mathbf{n})} e_\ell
\;-\;
\sum_{\ell\in\text{in}(\mathbf{n})} e_\ell
\;=\; 0
\quad
\forall \mathbf{n}.
\]

Hence, physically allowed flux patterns must form closed loops or wrap around periodic boundaries, ensuring no net sources or sinks. This arises from Gauss’s law, which mandates that the integer flux quanta $e_\ell$ on each link correspond to conserved flux lines. For systems with periodic boundary conditions, flux lines may topologically wind around the lattice in the $x$ or $y$ directions, defining distinct winding sectors. Thus, only closed-loop configurations or globally wrapped flux (without endpoints) are permitted.

\subsection{Truncating $\mathrm{U}(1)$ Gauge Fields}
\label{sec:U1_LinkVariables}

In \emph{compact} $\mathrm{U}(1)$ lattice gauge theory, each link $\ell$ carries two canonical operators:

\begin{enumerate}
	\item $\hat{\theta}_\ell$: the link phase (or discrete vector potential), taking values in the continuous interval $[0,2\pi)$.  
	\item $\hat{E}_\ell$: the integer-valued electric field operator, whose eigenvalues $e_\ell \in \mathbb{Z}$ correspond to flux quanta along link $\ell$.
\end{enumerate}

These satisfy commutation relations similar to the canonical pair $\bigl[\hat{\theta}_\ell, \hat{E}_{\ell'}\bigr] = i\,\delta_{\ell,\ell'}$.  In the continuum analog, $\hat{E}_\ell $ corresponds to the derivative operator $-i\,\frac{\partial}{\partial \theta_\ell}$.

\paragraph{Truncation to Spin-$S$ (Quantum Link Models).}
In the traditional Kogut-Susskind formulation, $\hat{E}_\ell$ can take any integer value, so the Hilbert space on each link is infinite-dimensional.  However, for practical digital quantum simulations, it is often useful to truncate the electric field to a bounded range $-S \le e_\ell \le +S$.  One then identifies:
\[
\hat{E}_\ell \;\longmapsto\; S^z_\ell, 
\quad
\hat{U}_\ell = e^{i\hat{\theta}_\ell} \;\longmapsto\; S^+_\ell,
\]
where $S^z_\ell$ and $S^\pm_\ell$ are spin-$S$ operators acting in a $(2S+1)$-dimensional Hilbert space (qudits).  This finite-dimensional \emph{quantum link model} preserves local gauge invariance while restricting large flux excitations.  

Throughout theoretical discussions, one may keep the full infinite-dimensional space for conceptual clarity.  In practical quantum hardware implementations, however, the truncated spin-$S$ version is more suitable, mapping each link to a finite register of qubits or qudits.

\subsection{Trotterized Simulation Strategies for $\mathrm{U}(1)$}
\label{sec:U1_Trotter}

Digital quantum simulators often approximate real-time evolution of the pure $\mathrm{U}(1)$ Hamiltonian using a Suzuki-Trotter decomposition.  By separating the Hamiltonian $\hat{H}$ into electric and magnetic parts,
\[
\hat{H}_E 
\;=\; 
\frac{g^2}{2}\,\sum_{\ell} \hat{E}_\ell^2,
\quad
\hat{H}_B 
\;=\; 
-\;\frac{1}{2g^2}\,\sum_{\Box}\bigl(\hat{U}_\Box + \hat{U}_\Box^\dagger\bigr),
\]
the time-evolution operator can be approximated as
\[
e^{-i\hat{H}t}
\;\approx\;
\Bigl[e^{-i\hat{H}_E\,\Delta t}\;e^{-i\hat{H}_B\,\Delta t}\Bigr]^{t/\Delta t}.
\]

\paragraph{Electric Step.}
The ``electric'' step $e^{-i\hat{H}_E\,\Delta t}$ is diagonal in the flux basis $\ket{e_\ell}$.  Each link $\ell$ simply acquires a phase 
$\exp\!\bigl[-\,i\,\tfrac{g^2}{2}\,e_\ell^2\,\Delta t\bigr]$.  

\paragraph{Magnetic Step.}
The ``magnetic'' step $e^{-i\hat{H}_B\,\Delta t}$ couples different flux configurations through the plaquette operators $\hat{U}_\Box$.  In practice, local gates acting on each plaquette (or subsets of plaquettes) are applied sequentially to update flux values, while preserving gauge invariance.  Since $\hat{U}_\Box$ commutes with the gauge constraints $\hat{G}_{\mathbf{n}}$, if the initial state lies in the physical subspace, it remains there (aside from Trotter errors).

By studying how observables evolve under Trotterized steps, one can investigate various phases of the $\mathrm{U}(1)$ model - from strong coupling (where flux lines localize) to regimes in which quantum fluctuations delocalize or cause flux loops to ``resonate.''  Crucially, gauge invariance is maintained at every step, so the dynamics remain in the physical subspace, and measurements always reflect physically valid, packaged excitations.

\subsection{Preparing and Measuring Packaged Quantum States for $\mathrm{U}(1)$}

In a pure $\mathrm{U}(1)$ lattice gauge theory (i.e., without matter fields), integer-valued electric fluxes on the links constitute the fundamental excitations.  Gauss’s law enforces that each site has zero net flux, so flux lines can only form closed loops or wrap around the torus.  This section illustrates how the \emph{packaging principle} appears in such Abelian models, and how one can prepare and measure the resulting flux-loop states.

\paragraph{Integer Flux and the Packaging Principle.}
Each link $\ell$ carries an electric field operator $\hat{E}_\ell$ with integer eigenvalues $e_\ell \in \mathbb{Z}$, representing the irreps of $\mathrm{U}(1)$.  Hence, one cannot create ``half'' a flux quantum or distribute flux fractionally across multiple links.  Because there are no matter fields, Gauss’s law at every site
\[
\sum_{\ell\in \text{out}(\mathbf{n})} e_\ell
\;-\;
\sum_{\ell\in \text{in}(\mathbf{n})} e_\ell
= 0
\]
forces flux lines to close on themselves, forming loops or winding around the periodic lattice. In this sense, flux quanta are packaged as indivisible units.

\paragraph{Flux-Loop States.}
A simple flux-loop state is created by assigning an integer flux $e$ to all links in a closed contour $\mathcal{C}$, while setting the flux on other links to zero:
\[
\ket{\Psi_{\mathrm{loop}}} 
\;=\;
\bigotimes_{\ell\in \mathcal{C}} \ket{e_\ell = e}
\; \otimes \;
\bigotimes_{\ell\notin \mathcal{C}} \ket{e_\ell = 0}.
\]
Since $\mathcal{C}$ is a closed loop, Gauss’s law is satisfied at each vertex.  More complex loop states arise by superimposing multiple loops or assigning different integer flux values to different loops.

\paragraph{Winding Superselection.}
On a 2D torus, one defines global winding numbers $W_x$ and $W_y$ that measure total flux crossing each noncontractible direction.  Because local gauge-invariant operations cannot cut a flux line that encircles the lattice, each pair $(W_x, W_y)$ identifies a distinct topological sector.  Cross-sector superpositions, such as $\alpha\ket{W_x=0} + \beta\ket{W_x=1}$, are unphysical under purely local dynamics, thereby reflecting a superselection rule associated with global flux.

\paragraph{Superpositions and Resonating Flux Loops.}
Within a single winding sector, nontrivial packaged entanglement arises from superpositions of different loop configurations.  On a 2D torus, for instance,
\[
\ket{\Psi}
\;=\;
\alpha\,\ket{\text{Loop}_x} 
\;+\;
\beta\,\ket{\text{Loop}_y},
\quad
|\alpha|^2 + |\beta|^2=1,
\]
combines an $x$-direction flux loop with a $y$-direction loop.  Although each loop state alone is gauge-invariant, their superposition creates  packaged entanglement across multiple links.  In general, a magnetic Hamiltonian term $\hat{U}_\Box$ can flip flux around plaquettes, generating a ``resonance'' among multiple loop coverings and enhancing many-body packaged entanglement.

\paragraph{Preparing Flux-Loop States.}
Such states can be engineered on a quantum simulator via:
\begin{itemize}
	\item \emph{Initial Configuration:}  
	Initialize a simple product state (e.g., a single flux loop).
	\item \emph{Adiabatic Ramping of Plaquette Terms:}  
	Slowly increase the terms allowing flux to hop or flip around plaquettes.  If done slowly relative to the energy gaps, the state can evolve into a resonating flux-loop superposition.
	\item \emph{Gate-by-Gate Methods (Digital Simulation):}  
	Construct local $\mathrm{U}(1)$-covariant gates (e.g., link flip or controlled phase) that create specific loop superpositions.
\end{itemize}

\paragraph{Gauge-Invariant Measurements.}
After evolving under Trotterized or adiabatic protocols, relevant observables include:
\begin{itemize}
	\item \textbf{Electric Energy:}
	$\sum_{\ell} \langle \hat{E}_\ell^2 \rangle$, probing overall flux magnitude.
	\item \textbf{Magnetic Plaquette:}
	$\langle \hat{U}_\Box \rangle$ or $\langle \hat{U}_\Box + \hat{U}_\Box^\dagger\rangle$, indicating local flux alignment around plaquettes.
	\item \textbf{Wilson Loops:}
	$W(\mathcal{C}) = \mathrm{Re}\,\langle \prod_{\ell \in \mathcal{C}} \hat{U}_\ell \rangle$, used to diagnose confinement-like behavior or flux patterning.
	\item \textbf{Winding Numbers:}
	$\langle \hat{W}_x \rangle$ or $\langle \hat{W}_y \rangle$, detecting net flux around the torus.
\end{itemize}

One can also measure link operators $\hat{E}_\ell$ directly to infer flux distributions, or use partial tomography (feasible in smaller systems) to quantify packaged entanglement. Throughout, the integer flux constraint and Gauss’s law guarantee that only closed, gauge-invariant flux excitations appear in the physical subspace.

\section{$\mathrm{SU}(2)$ Lattice Gauge Theory}
\label{sec:SU2Gauge}

We now extend our lattice gauge theory (LGT) framework from the Abelian $\mathrm{U}(1)$ group (Section~\ref{sec:U1Gauge}) to the non-Abelian gauge group $\mathrm{SU}(2)$.\cite{Yang1954,Wilson1974,KogutSusskind1975,Zohar2011,ZoharPRL2013,Creutz1980}
While many foundational ideas (Gauss’s law, link Hilbert spaces, Trotterization) carry over, new features arise from the matrix-valued nature of non-Abelian transformations and the richer structure of color flux irreps (e.g.\ spin-$\tfrac12$, spin-1).  
In this section, we illustrate how $\mathrm{SU}(2)$ lattice gauge theory enforces the \emph{packaging principle} - no partial or fractional color - and how this underlies phenomena like color confinement and flux-tube formation.

\subsection{2D Lattice Setup for $\mathrm{SU}(2)$}
\label{sec:SU2_lattice}

As before, consider a 2D square lattice of size $L \times L$ with periodic boundary conditions.  Each site $\mathbf{n} = (n_x,n_y)$ has two outgoing links: one in the $\hat{x}$ direction and one in the $\hat{y}$ direction, for a total of $2L^2$ links.  Now, each link $\ell$ carries an $\mathrm{SU}(2)$ gauge field.  

The choice of 2D again helps us illustrate flux constraints in a simpler geometry - though all ideas extend to 3D or 4D.  $\mathrm{SU}(2)$ historically arises in Yang-Mills theory \cite{Yang1954} and is a stepping stone toward more complex non-Abelian groups like $\mathrm{SU}(3)$.

\subsection{Kogut-Susskind Hamiltonian for Pure $\mathrm{SU}(2)$}
\label{sec:SU2_Hamiltonian}

In close analogy to the $\mathrm{U}(1)$ case, the standard Kogut-Susskind Hamiltonian for a pure $\mathrm{SU}(2)$ gauge theory (in either 2+1 or 3+1 dimensions) is given by (see Eq.(\ref{GaugeTheoryHamiltonians}))
\begin{equation}
	\hat{H}
	\;=\;
	\underbrace{\frac{g^2}{2}\sum_{\ell} \hat{\mathbf{E}}_\ell^2}_{\text{electric term}}
	\;+\;
	\underbrace{\frac{1}{2g^2}\,\sum_{\Box}
		\mathrm{Tr}\Bigl(\hat{U}_\Box + \hat{U}_\Box^\dagger\Bigr)}_{\text{magnetic (plaquette) term}},
	\label{eq:SU2Hamiltonian}
\end{equation}
where:
\begin{itemize}
	\item $\hat{\mathbf{E}}_\ell^2 = \sum_{a=1}^{3} \bigl(\hat{E}_\ell^a\bigr)^2$ is the quadratic Casimir of the local color-electric field (in the adjoint representation), which penalizes large color flux.
	\item $\hat{U}_\Box$ is the product of link operators around an elementary plaquette $\Box$.  Concretely,
	\[
	\hat{U}_\Box
	\;=\;
	\hat{U}_{(\mathbf{n},\hat{\mu})}\,
	\hat{U}_{(\mathbf{n}+\hat{\mu},\hat{\nu})}\,
	\hat{U}^\dagger_{(\mathbf{n}+\hat{\nu},\hat{\mu})}\,
	\hat{U}^\dagger_{(\mathbf{n},\hat{\nu})},
	\]
	and the trace $\mathrm{Tr}(\cdot)$ is taken in the fundamental representation of $\mathrm{SU}(2)$.
\end{itemize}

Both the electric term and the magnetic (plaquette) term commute with the local gauge transformation generators $\hat{G}_{\mathbf{n}}^a$.  Consequently, the time evolution governed by $\hat{H}$ remains within the physical (gauge-invariant) subspace $\mathcal{H}_{\mathrm{phys}}$.

\paragraph{Quantum Link Truncation.}
In many \emph{quantum link model} implementations, the operators $\hat{E}_\ell^a$ and $\hat{U}_\ell$ are replaced by suitable spin-$\mathbf{S}_\ell$ operators of finite dimension $2S+1$, preserving the $\mathrm{SU}(2)$ algebra up to truncation.  This allows a finite-dimensional representation of each link, making the model amenable to digital or analog quantum simulation.

\subsection{Local $\mathrm{SU}(2)$ Gauge Transformations and Gauss’s Law}
\label{sec:SU2_gausslaw}

At each site $\mathbf{n}$, a local gauge transformation is specified by a unitary matrix $g_{\mathbf{n}} \in \mathrm{SU}(2)$.  This transformation acts on the link variables as follows:
\[
\hat{U}_{(\mathbf{n},\hat{\mu})}
\;\longmapsto\;
g_{\mathbf{n}}\;\hat{U}_{(\mathbf{n},\hat{\mu})},
\quad
\hat{U}_{(\mathbf{n}-\hat{\mu},\hat{\mu})}
\;\longmapsto\;
\hat{U}_{(\mathbf{n}-\hat{\mu},\hat{\mu})}\;g_{\mathbf{n}}^\dagger,
\]
where $\hat{U}_{(\mathbf{n},\hat{\mu})}$ is the link operator emanating from site $\mathbf{n}$ in direction $\hat{\mu}$, and $\hat{U}_{(\mathbf{n}-\hat{\mu},\hat{\mu})}$ is the link operator entering $\mathbf{n}$ from $\mathbf{n}-\hat{\mu}$.  Simultaneously, the color-electric field operators $\hat{E}_\ell^a$ transform in the \emph{adjoint} representation of $\mathrm{SU}(2)$.  By design, these local transformations leave the plaquette operators $\hat{U}_\Box$ gauge-invariant.

\paragraph{Gauss’s Law in Pure $\mathrm{SU}(2)$.}
In a \emph{pure} gauge theory with no matter fields, Gauss’s law requires zero net color flux at each site.  Concretely, defining
\[
\hat{G}_{\mathbf{n}}^a
\;=\;
\sum_{\ell \in \text{in}(\mathbf{n})} \hat{E}_\ell^a 
\;-\;
\sum_{\ell \in \text{out}(\mathbf{n})} \hat{E}_\ell^a,
\]
the physical states $\ket{\Psi_{\mathrm{phys}}}$ must satisfy
\begin{equation}\label{Eq:SU2GaussLaw}
	\hat{G}_{\mathbf{n}}^a \,\ket{\Psi_{\mathrm{phys}}} 
	\;=\;
	0
	\quad
	\text{for all}\;\mathbf{n},\;\;a=1,2,3.
\end{equation}
This implies color flux lines must form closed loops or vanish entirely, mirroring the ``no net flux'' condition in $\mathrm{U}(1)$, but now in a non-Abelian context.

\paragraph{Including Matter Fields.}
If matter in the fundamental (doublet) representation is present at site $\mathbf{n}$, Gauss’s law generalizes to
\[
\sum_{\ell\in \text{in}(\mathbf{n})}\hat{E}_\ell^a
\;-\;
\sum_{\ell\in \text{out}(\mathbf{n})}\hat{E}_\ell^a
\;=\;
\hat{T}_{\mathbf{n}}^a,
\]
where $\hat{T}_{\mathbf{n}}^a$ are the color generators acting on the quark (or matter) field.  Hence, the total flux plus the matter’s color charge at each site must vanish, ensuring local color neutrality (singlet formation).

\subsection{Physical (Gauge-Invariant) Hilbert Space}
\label{sec:SU2_physical}

As in the $\mathrm{U}(1)$ case, the total unconstrained Hilbert space for an $\mathrm{SU}(2)$ lattice gauge theory factorizes over all links:
\[
\mathcal{H}
\;=\;
\bigotimes_{\ell=1}^{2L^2}
\mathcal{H}_\ell^{(\mathrm{link})},
\]
where each $\mathcal{H}_\ell^{(\mathrm{link})}$ is an $\mathrm{SU}(2)$ representation space (e.g., spin-$S$ if truncated).  However, physical states must satisfy \emph{Gauss’s law} at every site $\mathbf{n}$:
\[
\hat{G}_{\mathbf{n}}^a\,\ket{\Psi}
\;=\;
0,
\quad 
\forall\,\mathbf{n},\,a=1,2,3,
\]
where $\hat{G}_{\mathbf{n}}^a$ enforces the local color neutrality (no net color flux).  Thus, the \textbf{physical subspace} $\mathcal{H}_{\mathrm{phys}}$ is defined by
\[
\mathcal{H}_{\mathrm{phys}}
\;=\;
\Bigl\{
\ket{\Psi}\in \mathcal{H}
\;\Big\vert\;
\hat{G}_{\mathbf{n}}^a\,\ket{\Psi}=0,\;\forall\,\mathbf{n},a
\Bigr\}.
\]
Equivalently, one can view $\ket{\Psi}$ in $\mathcal{H}_{\mathrm{phys}}$ as invariant under local $\mathrm{SU}(2)$ gauge transformations.

\paragraph{Packaged (Color Flux) Interpretation.}
\begin{itemize}
	\item In $\mathrm{SU}(2)$, the flux on each link can be labeled by a spin-$j$ irrep.  
	\item Gauss’s law requires that the total color flux at each site sums to zero, implying flux lines either close on themselves or attach to matter fields.  
	\item Crucially, ``partial color'' is forbidden - a link’s flux must be a full spin-$j$ representation, reinforcing the \textbf{packaging principle}: no fractional color can appear.
\end{itemize}

Hence, in the absence of matter, $\mathrm{SU}(2)$ flux lines must form closed, non-Abelian loops or be entirely absent; if matter is present, it carries fundamental color charge that must combine with flux to form an overall color singlet at each site.

\subsection{Truncating $\mathrm{SU}(2)$ Gauge Fields}
\label{sec:SU2_linkspace}

In the Kogut-Susskind formulation of $\mathrm{SU}(2)$ lattice gauge theory, each link $\ell$ carries two sets of operators:

\paragraph{1. Link Operator $\hat{U}_\ell$.}
$\hat{U}_\ell$ is a $2\times 2$ special unitary matrix operator ($\mathrm{SU}(2)$), serving as the parallel transporter between two adjacent lattice sites.  It generalizes the Abelian phase operator $\hat{U}_\ell = e^{i\hat{\theta}_\ell}$ to a non-Abelian context, where phases become matrix degrees of freedom.

\paragraph{2. Color-Electric Fields $\hat{E}_\ell^a$.}
These three components ($a=1,2,3$) generate the local $\mathfrak{su}(2)$ Lie algebra on link $\ell$, with commutation relations
\[
[\,\hat{E}_\ell^a,\,\hat{E}_\ell^b\,] 
\;=\; 
i\,\epsilon^{abc}\,\hat{E}_\ell^c,
\quad
[\,\hat{E}_\ell^a,\,\hat{U}_\ell\,]\;\neq\;0.
\]
Physically, $\hat{E}_\ell^a$ can be viewed as a ``color spin'' measuring non-Abelian electric flux along the link.

\paragraph{Infinite-Dimensional vs. Quantum Link Truncation.}
\begin{itemize}
	\item \textbf{Classical Kogut-Susskind Formulation:}
	$\hat{U}_\ell$ spans the continuous $\mathrm{SU}(2)$ group, and the associated electric field $\hat{E}_\ell^a$ can be unbounded in magnitude.  This yields an infinite-dimensional link Hilbert space.
	
	\item \textbf{Quantum Link Model (QLM):}
	To make the theory more tractable for quantum simulation, one can truncate each link to a finite spin-$S$ representation, reducing the dimension to $2S+1$.  In this truncated picture, $\hat{\mathbf{E}}_\ell^2$ maps onto the spin Casimir $\hat{\mathbf{S}}_\ell^2$, and the operator $\hat{U}_\ell$ effectively behaves like spin raising/lowering operators (up to representation factors).  This approach retains local $\mathrm{SU}(2)$ gauge invariance while making the Hilbert space finite-dimensional - suitable for digital or analog quantum hardware.
\end{itemize}

Whether in the full (infinite-dimensional) or truncated (finite-dimensional) representation, the link Hilbert space $\mathcal{H}_\ell$ carries a non-Abelian $\mathrm{SU}(2)$ structure, with $\hat{U}_\ell$ and $\hat{E}_\ell^a$ obeying the canonical commutation relations of the gauge group.

\subsection{Adding Matter Fields}

If we add matter in the fundamental representation $\mathbf{2}$ of $SU(2)$ (like ``quarks'' for an $SU(2)$ toy model), each site $\mathbf{n}$ has a matter field operator $\hat{\psi}_{\mathbf{n},\alpha}$ with $\alpha\in\{1,2\}$.  Then Gauss’s law modifies to
\[
\sum_{\ell\in\text{in}(\mathbf{n})}
\hat{E}_\ell^a
\;-\;
\sum_{\ell\in\text{out}(\mathbf{n})}
\hat{E}_\ell^a
\;=\;
\hat{T}^a(\mathbf{n})\,,
\]
where $\hat{T}^a(\mathbf{n})$ are the color generators for the matter field in the fundamental representation.  Thus, if you place a single ``quark'' (spin-1/2 in color space) at site $\mathbf{n}$, the outgoing color flux must cancel its $\mathbf{2}$ representation so the entire site is color-singlet overall.

According to the idea of packaging, a quark in representation $\mathbf{2}$ plus a link flux in representation $\overline{\mathbf{2}}$ can form a singlet, or multiple quarks can form higher representations, but no partial color can appear. This extends the irreps logic from single-particle excitations to multi-particle color singlets - reinforcing that color cannot be fractionally assigned across different excitations.

\subsection{Trotterized Simulation Strategies for $\mathrm{SU}(2)$}
\label{sec:SU2_Trotter}

A common approach to simulating real-time dynamics of a pure $\mathrm{SU}(2)$ gauge theory on digital quantum hardware is to split the Hamiltonian into electric and magnetic parts and apply a Suzuki-Trotter decomposition.  Specifically, one writes
\[
\hat{H}
\;=\;
\hat{H}_E + \hat{H}_B,
\quad
\text{where}
\quad
\hat{H}_E 
= \frac{g^2}{2}\sum_\ell \hat{\mathbf{E}}_\ell^2,
\quad
\hat{H}_B 
= \frac{1}{2g^2}\sum_{\Box} \mathrm{Tr}\bigl(\hat{U}_\Box + \hat{U}_\Box^\dagger\bigr).
\]
The time-evolution operator is then approximated by
\[
e^{-\,i\,\hat{H}\,\Delta t}
\;\approx\;
e^{-\,i\,\hat{H}_E\,\Delta t}
\;e^{-\,i\,\hat{H}_B\,\Delta t},
\]
repeated $t/\Delta t$ times to cover total evolution $t$.

\paragraph{Electric and Magnetic Steps.}
\begin{itemize}
	\item \emph{Electric Step:} $e^{-\,i\,\hat{H}_E\,\Delta t}$ is diagonal in the basis labeling each link’s color flux (e.g., spin-$j$ in a quantum link model).  It applies phases $\exp\!\bigl[-\,i\,(g^2/2)\,\hat{\mathbf{E}}_\ell^2\,\Delta t\bigr]$ to each link, penalizing large color flux.
	\item \emph{Magnetic Step:} $e^{-\,i\,\hat{H}_B\,\Delta t}$ couples neighboring links through $\hat{U}_\Box$ operators, ``rotating'' or ``flipping'' color flux around plaquettes.  In a finite spin-$S$ representation, these become local spin-exchange operations.
\end{itemize}
Because $\hat{\mathbf{E}}_\ell^2$ and $\mathrm{Tr}\bigl(\hat{U}_\Box + \hat{U}_\Box^\dagger\bigr)$ commute with the gauge generators $\hat{G}_{\mathbf{n}}^a$, each Trotter step preserves gauge invariance (aside from small Trotter errors).  Consequently, the evolving state remains in the physical subspace $\mathcal{H}_{\mathrm{phys}}$.

\paragraph{Gauge-Invariant Observables.}
Once the system has evolved via Trotterized steps, one can measure observables that probe confinement and non-Abelian flux dynamics:
\begin{itemize}
	\item \textbf{Electric Energy:} 
	$\sum_\ell \langle \hat{\mathbf{E}}_\ell^2\rangle$, indicating color flux strength on each link.
	\item \textbf{Magnetic Plaquette:} 
	$\langle \mathrm{Tr}\,\hat{U}_\Box\rangle$ or $\langle \mathrm{Tr}(\hat{U}_\Box + \hat{U}_\Box^\dagger)\rangle$, revealing how flux is distributed around elementary squares.
	\item \textbf{Wilson Loops:} 
	\[
	W(\mathcal{C})
	\;=\;
	\Bigl\langle \mathrm{Tr}\,\prod_{\ell\in \mathcal{C}} \hat{U}_\ell \Bigr\rangle,
	\]
	used to detect confinement via area-law vs.\ perimeter-law scaling for large loops $\mathcal{C}$.
\end{itemize}

\paragraph{Non-Abelian Flux Loops and Spin-$j$ Truncation.}
In a quantum link model for $\mathrm{SU}(2)$, each link is truncated to a finite spin-$S$ representation.  The eigenvalues of $\hat{\mathbf{E}}_\ell^2$ are labeled by an angular momentum quantum number $j \leq S$.  Gauss’s law at each vertex enforces that the \emph{vector sum} of incoming and outgoing color fluxes vanishes, implying flux lines must form closed loops or attach to matter in a color-neutral manner.  One can construct an excited ``flux-loop'' state by assigning nonzero spin-$j$ values along a closed contour of links, then superposing such configurations to form resonating flux loops. This exemplifies a multi-particle packaged entangled state: each link carries a full irreps (spin-$j$), and all links together satisfy local color neutrality.

\subsection{Preparing and Measuring Packaged Quantum States for $\mathrm{SU}(2)$}
\label{sec:SU2_packaging}

$\mathrm{SU}(2)$ gauge invariance enforces that each link carries a full spin-$j$ representation (no partial color flux), while Gauss’s law requires the total color flux at each lattice site to vanish (or to match any matter fields present).  These rules reflect the \emph{packaging principle}: each non-Abelian flux line is a complete irreps block, either closing on itself or forming color-neutral combinations with matter.

\paragraph{Non-Abelian Packaging: Closed Color Flux and Irrep Matching.}
\begin{enumerate}
	\item \textbf{Link Irreps (Spin-$j$):}  
	Each link is in a spin-$j$ representation of $\mathrm{SU}(2)$.  One cannot split, say, spin-$\tfrac12$ across multiple partial excitations.  Valid values are $j=0,\tfrac12,1,\tfrac32,\dots$, depending on whether one adopts an infinite or truncated Hilbert space.
	\item \textbf{No Partial Color:}  
	Gauss’s law $\hat{G}_{\mathbf{n}}^a=0$ implies that the total color flux (a vector in $\mathfrak{su}(2)$) at each site must be zero, leading flux lines to either form closed loops or connect matter fields (fundamental $\mathbf{2}$, antiquark $\overline{\mathbf{2}}$, etc.) into color singlets.
	\item \textbf{Color Singlets with Matter:}  
	If quarks (in $\mathbf{2}$) and antiquarks (in $\overline{\mathbf{2}}$) are introduced, the net color at each site must still vanish, so flux plus matter forms a local singlet $\mathbf{1}$.  This encapsulates ``no partial color,'' aligning with the idea of color confinement.
\end{enumerate}

\paragraph{Example: Mesons and Flux-Tube Superpositions.}
When fundamental matter is present, a canonical illustration is the formation of \emph{meson} states:
\begin{itemize}
	\item A quark in representation $\mathbf{2}$ at site $\mathbf{n}$,
	\item An antiquark in $\overline{\mathbf{2}}$ at site $\mathbf{m}$.
\end{itemize}
Gauss’s law forces a color flux tube connecting these two sites into a total singlet $\mathbf{2}\otimes\overline{\mathbf{2}}\to\mathbf{1}$.  If multiple lattice paths exist, one can form superpositions:
\[
\ket{\Phi}
\;=\;
\alpha\,\ket{\text{Tube}_1}
\;+\;
\beta\,\ket{\text{Tube}_2},
\]
where each $\ket{\text{Tube}_i}$ is gauge-invariant on its own (one flux tube from $\mathbf{n}$ to $\mathbf{m}$), but their superposition creates packaged entanglement across different flux-tube routes. Plaquette interactions ($\mathrm{Tr}\,\hat{U}_\Box$) can further cause these flux tubes \cite{Bali1995,Bali2001,Takahashi2002} to shift or ``resonate,'' reminiscent of the $\mathrm{U}(1)$ resonating-loop scenario but with richer non-Abelian color dynamics.

\paragraph{Flux Loops in Pure Gauge Theories.}
Without matter, flux lines must close on themselves.  In spin-$j$ quantum link models, one can construct closed-loop excitations by assigning nonzero $j$ along a loop and zero elsewhere, ensuring color neutrality at each vertex.  Superpositions of such closed loops yield nontrivial multi-link packaged entanglement while respecting $\mathrm{SU}(2)$ invariance.

\paragraph{Packaged Entanglement and Measurement.}
To probe these packaged states, one can measure:
\begin{itemize}
	\item \textbf{Link Electric Fields:} $\langle\hat{\mathbf{E}}_\ell^2\rangle$ to quantify flux magnitudes.
	\item \textbf{Plaquette Operators:} $\langle \mathrm{Tr}\,\hat{U}_\Box\rangle$ revealing flux orientation and ``resonance'' among loops or tubes.
	\item \textbf{Wilson Loops:} $\mathrm{Tr}\,\prod_{\ell\in\mathcal{C}}\hat{U}_\ell$ to check for confinement (area law) or screening effects.
	\item \textbf{Packaged Entanglement Entropy:} Partitioning the lattice into regions and computing (or bounding) packaged entanglement entropies can show how color flux lines spread across different subsystems.
\end{itemize}
In all cases, \emph{packaging} ensures that color excitations appear only as full irreps, and Gauss’s law enforces local color neutrality.  This fundamentally shapes the structure of gauge-invariant multi-particle states - whether mesons, baryons, or flux loops - and drives essential phenomena like confinement and hadron formation in $\mathrm{SU}(2)$.

\section{$\mathrm{SU}(3)$ Lattice Gauge Theory}
\label{sec:SU3Gauge}

We now move from $\mathrm{SU}(2)$ to $\mathrm{SU}(3)$, the gauge group of Quantum Chromodynamics (QCD) \cite{Wilson1974,KogutSusskind1975,Creutz1981}.  In high-energy physics, $\mathrm{SU}(3)$ describes the color interactions of quarks and gluons.  Consequently, the \textbf{packaging principle} (no partial or fractional color) is directly linked to color confinement: physical hadrons must be color-singlet states.  We show how local $\mathrm{SU}(3)$ transformations, Gauss’s law, and the standard Kogut-Susskind Hamiltonian enforce the same fundamental constraints as in $\mathrm{U}(1)$ or $\mathrm{SU}(2)$, but now with richer color flux lines ($\mathbf{3}$, $\overline{\mathbf{3}}$, $\mathbf{8}$, etc.).

\subsection{Lattice Setup and $\mathrm{SU}(3)$ Link Variables}
\label{sec:SU3_lattice}

As in previous sections, consider a 2D (or 3D) square lattice of linear size $L$ and periodic boundaries:
\[
\mathbf{n} = (n_x,n_y), \quad n_x,n_y \in \{0,\dots,L-1\}.
\]
Each site $\mathbf{n}$ has links in the $+x$ and $+y$ directions (2D case), for a total of $2L^2$ links in 2D.  Now, each link $\ell$ carries an $\mathrm{SU}(3)$ gauge field:

1. Link Operator $\hat{U}_\ell \in \mathrm{SU}(3)$:  
A $3\times3$ special unitary matrix operator, the non-Abelian generalization of the gauge link.  

2. Color-Electric Fields $\hat{E}_\ell^a$ ($a=1,\dots,8$):  
These form the $\mathfrak{su}(3)$ Lie algebra, satisfying
\[
[\,\hat{E}_\ell^a,\hat{E}_\ell^b\,] 
= i\,f^{abc}\,\hat{E}_\ell^c,
\quad
[\,\hat{E}_\ell^a,\hat{U}_\ell\,]\neq 0,
\]
where $f^{abc}$ are the $\mathrm{SU}(3)$ structure constants.  

In the pure Kogut-Susskind approach, $\hat{U}_\ell$ is continuous and $\hat{\mathbf{E}}_\ell^2$ can be unbounded.  For quantum simulation, one often \emph{truncates} to a finite subset of $\mathrm{SU}(3)$ irreps (a quantum link model), but the core gauge-invariant structure remains.

\subsection{Kogut-Susskind Hamiltonian for Pure $\mathrm{SU}(3)$}
\label{sec:SU3_hamiltonian}

In direct analogy to the $\mathrm{SU}(2)$ case, a pure $\mathrm{SU}(3)$ gauge theory in 2+1 or 3+1 dimensions is described by an electric term and a magnetic (plaquette) term (see Eq.(\ref{GaugeTheoryHamiltonians})):

\[
\hat{H}
\;=\;
\underbrace{\frac{g^2}{2}\sum_{\ell}\,\hat{\mathbf{E}}_\ell^2}_{\text{electric term}}
\;+\;
\underbrace{\frac{1}{2g^2}\sum_{\Box}\,\mathrm{Re}\,\mathrm{Tr}\bigl(\hat{U}_\Box\bigr)}_{\text{magnetic (plaquette) term}},
\]
where:
\begin{itemize}
	\item $\hat{\mathbf{E}}_\ell^2 = \sum_{a=1}^{8} \bigl(\hat{E}_\ell^a\bigr)^2$ is the quadratic Casimir of the color-electric field in the adjoint representation ($\mathbf{8}$), penalizing large color flux on link $\ell$.
	\item $\hat{U}_\Box$ is the product of $\mathrm{SU}(3)$ link operators around each plaquette $\Box$, and $\mathrm{Tr}$ is taken in the fundamental representation ($\mathbf{3}$). The operator $\mathrm{Re}\,\mathrm{Tr}\bigl(\hat{U}_\Box\bigr)$ drives magnetic flux fluctuations.
\end{itemize}

Both terms commute with local $\mathrm{SU}(3)$ gauge transformations, ensuring that time evolution remains in the physical, gauge-invariant subspace $\mathcal{H}_{\mathrm{phys}}$.  In practice, one may \emph{truncate} each link to a smaller set of $\mathrm{SU}(3)$ irreps (e.g., $\mathbf{3}, \overline{\mathbf{3}}, \mathbf{8}$, etc.) in a \emph{quantum link model} to achieve a finite-dimensional Hilbert space suitable for digital or analog quantum simulation.

\subsection{Local $\mathrm{SU}(3)$ Gauge Transformations and Gauss’s Law}
\label{sec:SU3_gausslaw}

A local gauge transformation at site $\mathbf{n}$ is specified by a matrix $g_{\mathbf{n}}\in \mathrm{SU}(3)$.  This transformation acts on the link operators as follows:
\[
\hat{U}_{(\mathbf{n},\hat{\mu})}
\;\longmapsto\;
g_{\mathbf{n}}\;\hat{U}_{(\mathbf{n},\hat{\mu})},
\quad
\hat{U}_{(\mathbf{n}-\hat{\mu},\hat{\mu})}
\;\longmapsto\;
\hat{U}_{(\mathbf{n}-\hat{\mu},\hat{\mu})}\;g_{\mathbf{n}}^\dagger,
\]
while the color-electric fields $\hat{E}_\ell^a$ (with $a=1,\dots,8$) transform in the adjoint representation.

\paragraph{Gauss’s Law in Pure $\mathrm{SU}(3)$.}
In a \emph{pure} gauge theory (i.e., no quarks), each site $\mathbf{n}$ must have zero net color flux:
\[
\hat{G}_{\mathbf{n}}^a 
\;=\;
\sum_{\ell\in \mathrm{in}(\mathbf{n})}\hat{E}_\ell^a
\;-\;
\sum_{\ell\in \mathrm{out}(\mathbf{n})}\hat{E}_\ell^a
\;=\;
0,
\quad a=1,\dots,8.
\]
Physical states $\ket{\Psi_{\mathrm{phys}}}$ satisfy
\begin{equation}\label{Eq:SU3GaussLaw}
	\hat{G}_{\mathbf{n}}^a\,\ket{\Psi_{\mathrm{phys}}}
	\;=\;
	0,
	\;\;\;
	\forall\,\mathbf{n},\,\forall\,a.
\end{equation}
Hence, $\mathrm{SU}(3)$ color flux lines can only form closed loops or wrap around the lattice boundaries, mirroring the closed-loop constraints in $\mathrm{U}(1)$ but with richer non-Abelian structure.

\paragraph{Inclusion of Quarks.}
If quarks in the fundamental representation $\mathbf{3}$ (and possibly antiquarks $\overline{\mathbf{3}}$) are introduced, Gauss’s law modifies to
\[
\sum_{\ell \in \mathrm{in}(\mathbf{n})}\hat{E}_\ell^a
\;-\;
\sum_{\ell \in \mathrm{out}(\mathbf{n})}\hat{E}_\ell^a
\;=\;
\hat{T}_{\mathbf{n}}^a,
\]
where $\hat{T}_{\mathbf{n}}^a$ are the color generators acting on the quark fields at site $\mathbf{n}$.  In other words, the net flux at each site combines with the local matter color charge to form a total singlet, prohibiting any partial or fractional color.

\begin{remark}
	Note that Gauss's law in Abelian case (see Eq.~\eqref{Eq:U1GaussLaw}) differs from that in non-Abelian cases (see Eqs.~\eqref{Eq:SU2GaussLaw} and \eqref{Eq:SU3GaussLaw}).
	This difference stems from the distinct structure of the gauge groups:
	\begin{itemize}
		\item \textbf{Abelian Case (U(1)):}  
		There is one generator at each lattice site. A local U(1) gauge transformation can be written in exponential form as
		\[
		\hat{G}_x = e^{i\alpha_x \hat{Q}_x},
		\]
		where \(\alpha_x\) is an arbitrary phase and \(\hat{Q}_x\) is the charge (or flux) operator. Invariance under such a transformation means that physical states are unchanged, i.e.,
		\[
		\hat{G}_x\,\ket{\Psi_{\mathrm{phys}}} = \ket{\Psi_{\mathrm{phys}}},
		\]
		which is equivalent to requiring that \(\hat{Q}_x\) has zero eigenvalue on \(\ket{\Psi_{\mathrm{phys}}}\). This single-condition neatly captures the absence of net charge (or flux) at each site.
		
		\item \textbf{Non-Abelian Case (SU(N)):}  
		Non-Abelian gauge groups possess multiple generators. Denote these by \(\hat{G}_x^a\) (with \(a = 1,2,\ldots, N^2-1\)). A local gauge transformation is expressed as
		\[
		\hat{U}_x(g) = \exp\Bigl(i\sum_a \alpha_x^a\,\hat{G}_x^a\Bigr).
		\]
		To ensure that a physical state is invariant under all such transformations, one must impose that
		\[
		\hat{G}_x^a\,\ket{\Psi_{\mathrm{phys}}} = 0\quad (\forall\, a).
		\]
		This condition means that the total ``non-Abelian charge'' or color flux at site \(x\) vanishes in every direction of the Lie algebra. In other words, the state forms a singlet under the local SU(N) transformation.
	\end{itemize}
\end{remark}

\subsection{Truncating $\mathrm{SU}(3)$ Gauge Fields}

In the Kogut-Susskind formulation of lattice $\mathrm{SU}(3)$, gauge-group variables and operators are as follows:

\begin{enumerate}
	\item  Link Operator $\hat{U}_\ell \in \mathrm{SU}(3)$.
	
	Each link is associated with a $3\times3$ special unitary matrix operator $\hat{U}_\ell$, representing the parallel transporter (or the discrete analog of the continuum gluon field $A_\mu$).  
	
	\item  Chromoelectric Field $\hat{E}_\ell^a$.
	
	A set of eight operators ($a=1,\dots,8$) generate local $\mathrm{SU}(3)$ transformations (the $\mathfrak{su}(3)$ Lie algebra).
	They obey nontrivial commutation relations,
	\[
	[\,\hat{E}_\ell^a,\,\hat{E}_\ell^b\,]
	= i\,f^{abc}\,\hat{E}_\ell^c,
	\quad
	[\,\hat{E}_\ell^a,\,\hat{U}_\ell\,]\neq 0,
	\]
	where $f^{abc}$ are the $\mathrm{SU}(3)$ structure constants.
\end{enumerate}

These link operators and color-electric fields transform under local $\mathrm{SU}(3)$ transformations, generalizing the notion of spin or integer flux from $\mathrm{U}(1)$ or $\mathrm{SU}(2)$ to a higher-dimensional gauge group.

For classical Kogut-Susskind model, $\hat{U}_\ell$ is a continuous $\mathrm{SU}(3)$ matrix; $\hat{E}_\ell^a$ can become unbounded as color flux operators.
For quantum link model (QLM), however, each link is truncated to a finite-dimensional representation of $\mathrm{SU}(3)$, e.g. restricting to certain highest-weight irreps. This yields a dimension suitable for qubit or qudit hardware, but must preserve the $\mathrm{SU}(3)$ algebra structure up to that cutoff.
In either approach, the link Hilbert space $\mathcal{H}_\ell$ is built from $\mathrm{SU}(3)$ irreps, guaranteeing that color excitations are packaged in complete representations.

\subsection{Adding Matter Fields in $\mathbf{3}$ or $\overline{\mathbf{3}}$}

To mimic quarks (fundamental color $\mathbf{3}$) or antiquarks ($\overline{\mathbf{3}}$) on the lattice, we place matter fields at each site. Then Gauss’s law becomes:

\[
\sum_{\ell\in\mathrm{in}(\mathbf{n})} \hat{E}_\ell^a
\;-\;
\sum_{\ell\in\mathrm{out}(\mathbf{n})} \hat{E}_\ell^a
=
\hat{T}^a(\mathbf{n}),
\]
where $\hat{T}^a(\mathbf{n})$ are the color generators for the matter field in representation $\mathbf{3}$ or $\overline{\mathbf{3}}$.

The net color flux at site $\mathbf{n}$ must exactly match the color charge of the matter occupant.  If a site has one quark in $\mathbf{3}$, the link flux must supply a $\overline{\mathbf{3}}$ to form a color singlet at that site overall. 

Hence, packaging arises because a single quark is an irreducible $\mathbf{3}$; partial color is disallowed, and you must have a matching $\overline{\mathbf{3}}$ flux (or multiple $\mathbf{3}$ quarks summing to color singlet).
Baryons (3 quarks: $\mathbf{3}\otimes\mathbf{3}\otimes\mathbf{3} \to \mathbf{1}$) or mesons (quark-antiquark: $\mathbf{3}\times\overline{\mathbf{3}}\to \mathbf{1}$) reflect packaged color singlets.

\subsection{Trotterized Simulation Strategies for $\mathrm{SU}(3)$}
\label{sec:SU3_trotter}

To simulate real-time dynamics of an $\mathrm{SU}(3)$ lattice gauge theory on a digital quantum device, one typically splits the Hamiltonian into electric and magnetic parts:

\[
\hat{H}
\;=\;
\hat{H}_E 
\;+\;
\hat{H}_B,
\quad
\text{where}
\quad
\hat{H}_E 
=\;
\frac{g^2}{2}\sum_\ell \hat{\mathbf{E}}_\ell^2,
\quad
\hat{H}_B
=\;
\frac{1}{2g^2}\sum_{\Box}\,\mathrm{Re}\,\mathrm{Tr}\!\bigl(\hat{U}_\Box\bigr).
\]
A Suzuki-Trotter decomposition approximates the time evolution as
\[
e^{-\,i\,\hat{H}\,\Delta t}
\;\approx\;
e^{-\,i\,\hat{H}_E\,\Delta t}
\;e^{-\,i\,\hat{H}_B\,\Delta t},
\]
repeated over small steps $\Delta t$.  Each sub-term acts on local subsets (links or plaquettes) and commutes with the gauge constraints $\hat{G}_{\mathbf{n}}^a$, so gauge invariance is preserved (up to Trotter errors).

\paragraph{Electric Step.}
$\hat{H}_E = \frac{g^2}{2}\sum_\ell \hat{\mathbf{E}}_\ell^2$ is diagonal in the color flux basis (or truncated spin basis if using a quantum link model).  In each Trotter step, this factor applies phases based on the squared color flux $\hat{\mathbf{E}}_\ell^2$.

\paragraph{Magnetic Step.}
$\hat{H}_B = \frac{1}{2g^2}\sum_{\Box}\mathrm{Re}\,\mathrm{Tr}(\hat{U}_\Box)$ couples link variables around each plaquette $\Box$.  It can ``flip'' or ``rotate'' color flux configurations, generating superpositions of different loop or tube states and capturing non-Abelian flux dynamics.

\paragraph{Gauge-Invariant Observables.}
Key measurements for probing $\mathrm{SU}(3)$ confinement and flux behavior include:
\begin{itemize}
	\item \textbf{Electric Energy:} 
	$\sum_{\ell}\langle \hat{\mathbf{E}}_\ell^2\rangle$, indicating how much color flux is present on each link.
	\item \textbf{Magnetic Plaquettes:} 
	$\langle\mathrm{Tr}(\hat{U}_\Box)\rangle$, revealing the distribution of flux around plaquettes.
	\item \textbf{Wilson Loops:} 
	\[
	W(\mathcal{C})
	\;=\;
	\Bigl\langle 
	\mathrm{Tr}\!\Bigl(
	\prod_{\ell\in \mathcal{C}} \hat{U}_\ell 
	\Bigr)
	\Bigr\rangle,
	\]
	testing whether flux tubes \cite{Bali1995,Bali2001,Takahashi2002} form between color sources (area law vs.\ perimeter law) and providing a hallmark signature of confinement in non-Abelian gauge theories.
	\item \textbf{Polyakov Loops (Finite Temperature) \cite{SvetitskyYaffe1982}:} 
	Measuring color deconfinement or screening in thermal ensembles.
	\item \textbf{Glueball/Hadron Correlation Functions (if matter is present):} 
	Revealing the mass spectra of bound states in $\mathrm{SU}(3)$.
\end{itemize}

By analyzing these observables under Trotterized evolution, one can investigate confinement, flux-tube formation, and non-Abelian phenomena, all while maintaining gauge invariance at the local level.

\subsection{Preparing and Measuring Packaged Quantum States for $\mathrm{SU}(3)$}
\label{sec:SU3_packaging}

In $\mathrm{SU}(3)$ gauge theories (as in QCD), color confinement and the requirement of \emph{color-singlet} physical states follow directly from local gauge invariance and the \textbf{packaging principle}.  Each quark (in $\mathbf{3}$), antiquark ($\overline{\mathbf{3}}$), or gluon ($\mathbf{8}$) is a complete irrep of $\mathrm{SU}(3)$.  Gauss’s law enforces that total color flux at each site must vanish or match matter color, ensuring that no ``partial color'' can exist as a separate excitation.  

\paragraph{1. No Partial Color: Confinement and Singlets.}
\begin{itemize}
	\item \textbf{Link Irreps:}  
	Gauge fields on each link can be truncated to certain $\mathrm{SU}(3)$ irreps (e.g., $\mathbf{1}$, $\mathbf{8}$, $\mathbf{3}$, or $\overline{\mathbf{3}}$) in quantum link models.  Each gluon or flux line is thus a full representation.
	\item \textbf{Local Gauss’s Law:}  
	At each site, the net color flux plus matter color must sum to zero (or to an overall singlet).  In pure gauge theory, flux lines form closed loops.  With matter, flux lines connect quark ($\mathbf{3}$) and antiquark ($\overline{\mathbf{3}}$) or bind three quarks into a baryon.
	\item \textbf{Color Superselection:}  
	States with different total color $\neq 0$ cannot be coherently superposed with color-singlet states, reflecting a superselection rule.  Only color-neutral excitations appear as physical asymptotic states, capturing the essence of confinement.
\end{itemize}

\paragraph{2. Composite Gauge-Invariant Excitations: Mesons, Baryons, Glueballs.}
\begin{itemize}
	\item \textbf{Mesons} $\bigl(\mathbf{3}\otimes\overline{\mathbf{3}}\to \mathbf{1}\bigr)$:  
	A quark and an antiquark form a color-singlet by contracting color indices (e.g., via the Kronecker delta). Multiple flux-tube paths connecting quark and antiquark can be superposed, creating packaged entangled states.
	\item \textbf{Baryons} $\bigl(\mathbf{3}\otimes\mathbf{3}\otimes\mathbf{3}\to \mathbf{1}\bigr)$:  
	Three quarks combine into a color singlet using the fully antisymmetric tensor $\epsilon_{\alpha\beta\gamma}$.  On the lattice, one may form multiple ``Y-shaped'' or ``star'' flux-tree configurations, again leading to superpositions.
	\item \textbf{Glueballs} $\bigl(\mathbf{8}\otimes\cdots\otimes\mathbf{8}\to \mathbf{1}\bigr)$:  
	Even in the pure gauge sector, color flux lines in the adjoint representation can bind into closed loops, forming glueball excitations.  These are color-singlet combinations of gluon fields.
\end{itemize}
All such states reflect \emph{multi-particle packaging}: each quark, antiquark, or gluon is a full $\mathrm{SU}(3)$ irrep, but their combination yields a net singlet under local color transformations.

\paragraph{3. Field Operators and Their Transformations.}
To illustrate quark/antiquark creation operators:
\[
\hat{\psi}^\dagger_{x,\alpha}:\quad \alpha=1,2,3 \quad (\mathbf{3}),
\qquad
\hat{\bar{\psi}}^\dagger_{x,\beta}:\quad \beta=1,2,3 \quad (\overline{\mathbf{3}}).
\]
Under a local gauge transformation $g_x \in \mathrm{SU}(3)$ at site $x$:
\[
\hat{\psi}^\dagger_{x,\alpha}
\;\longrightarrow\;
\sum_{\gamma}\,g_x(\alpha,\gamma)\,\hat{\psi}^\dagger_{x,\gamma},
\quad
\hat{\bar{\psi}}^\dagger_{x,\beta}
\;\longrightarrow\;
\sum_{\delta}\,\hat{\bar{\psi}}^\dagger_{x,\delta}\,\bigl(g_x^\dagger\bigr)(\delta,\beta).
\]
Thus, each operator is a full $\mathbf{3}$ or $\overline{\mathbf{3}}$ irreps, enforcing no fractional color excitations.

\paragraph{4. Meson and Baryon States: Concrete Constructions.}
\begin{itemize}
	\item \textbf{Meson:}
	\[
	\ket{M}
	\;=\;
	\sum_{\alpha=1}^{3}
	\hat{\psi}^\dagger_{x,\alpha}\,
	\hat{\bar{\psi}}^\dagger_{x,\alpha}\,\ket{0},
	\]
	contracting quark ($\mathbf{3}$) and antiquark ($\overline{\mathbf{3}}$) indices via the delta $\delta_{\alpha\alpha}$ to yield a color-singlet.  Under local $\mathrm{SU}(3)$, the unitarity of $g_x$ guarantees invariance.
	\item \textbf{Baryon:}
	\[
	\ket{B}
	\;=\;
	\epsilon_{\alpha\beta\gamma}\,
	\hat{\psi}^\dagger_{x,\alpha}\,
	\hat{\psi}^\dagger_{x,\beta}\,
	\hat{\psi}^\dagger_{x,\gamma}\,\ket{0},
	\]
	combining three quarks each in $\mathbf{3}$ into a totally antisymmetric color singlet using the Levi-Civita tensor $\epsilon_{\alpha\beta\gamma}$.  This is again invariant under local color transformations due to $\det(g_x)=1$.
\end{itemize}
Because each creation operator is packaged as a full irrep, these composite states automatically respect Gauss’s law, forbidding partial color at any site.

\paragraph{5. Preparing Packaged States on a Quantum Simulator.}
On a digital or analog quantum simulator:
\begin{itemize}
	\item \emph{Quark/Antiquark Register Encoding:}  
	A three-level (qutrit) system may encode quark color $\alpha \in \{0,1,2\}$.  Antiquarks are similarly encoded but transform under the conjugate representation.  
	\item \emph{Meson Construction Circuit:}  
	One can create a uniform superposition on the quark register, then apply a ``copy'' or controlled operation onto the antiquark register to form 
	$\frac{1}{\sqrt{3}} \sum_{\alpha} \ket{\alpha}_q \ket{\alpha}_{\bar{q}}$.  This reproduces the color-singlet structure.  
	\item \emph{Link Variables for Spatial Separation:}  
	For quark and antiquark at different sites, one inserts a Wilson line of gauge link operators between them.  On a digital device, this corresponds to a product of gauge-covariant gates acting along a path on the lattice, preserving gauge invariance.
\end{itemize}
All gates must respect local $\mathrm{SU}(3)$ transformations to remain in the physical subspace.

\paragraph{6. Measuring Packaged States: Observables and Confinement.}
After preparing mesons, baryons, or glueballs, gauge-invariant observables include:
\begin{itemize}
	\item \textbf{Wilson Loops:}
	$\mathrm{Tr}\!\bigl(\prod_{\ell\in\mathcal{C}}\hat{U}_\ell\bigr)$ tests whether color flux tubes remain confined (area law) or screen at large distances.
	\item \textbf{Meson/Baryon Correlation Functions:}
	Extracting mass gaps and hadron spectra.
	\item \textbf{Polyakov Loops (Finite Temperature) \cite{SvetitskyYaffe1982}:}
	Diagnosing color deconfinement transitions.
	\item \textbf{Packaged Entanglement Entropy:}
	Partitioning the lattice degrees of freedom to quantify how color flux lines entangle quarks/gluons across different regions.
\end{itemize}
Because color excitations remain packaged in full irreps, partial color excitations lie outside the physical subspace.  This aspect aids \emph{error mitigation}: any gauge-violating error that tries to produce fractional color is easily detected as leakage from the gauge-invariant sector.

\section{Discussion}

In this work, we have introduced the concept of packaged quantum states as a systematic approach to enforcing local gauge invariance in quantum simulations of lattice gauge theories. By encoding every link variable and matter excitation as a complete irrep of the gauge group, our method automatically enforces Gauss’s law and forbids the appearance of partial charges or colors. As a result, the physical Hilbert space is confined to gauge-invariant states from the outset - eliminating the need for large energy penalties or explicit gauge-fixing procedures.

This packaged-state approach provides several distinct advantages. First, it establishes an operational framework in which every lattice site or link is assigned a (possibly truncated) representation, ensuring that time evolution via Trotter steps preserves gauge constraints. Second, it offers a clear physical picture: for example, in $\mathrm{U}(1)$ theories the excitations manifest as quantized flux loops, while in non-Abelian theories (such as $\mathrm{SU}(2)$ or $\mathrm{SU}(3)$) color flux must form closed loops or bind to matter, reflecting confinement. Third, many gauge-violating processes are rendered either impossible or highly suppressed, which can reduce both the hardware requirements and the complexity of quantum circuits.

It is instructive to compare our method with two commonly employed strategies in lattice gauge theory simulations.
Table~\ref{tab:comparison_methods} summarizes the main pros and cons of each strategy:
\begin{enumerate}
	\item Gauge Fixing \cite{Rothe2012,Kogut1975,Zohar2015,Carena2024,Gribov1978}:
	By choosing a specific gauge (e.g., axial or Coulomb gauge), one reduces redundant degrees of freedom \cite{Rothe2012,Kogut1975}). This can simplify the Hilbert space and hardware requirements \cite{Tagliacozzo2013,Zohar2015}. However, gauge fixing may leave behind residual global constraints and is susceptible to ambiguities (e.g., Gribov copies) in non-Abelian theories.
%
%
	
	By contrast, the packaged approach never picks a gauge; it keeps the local symmetry manifest but simply restricts the physical Hilbert space to gauge‐invariant irreps from the start.
	
	\item Penalty Terms \cite{Banerjee2013,Hauke2013,Halimeh2020,ZoharPRA2013,Halimeh2022}:
	An additional energy term $\hat{H}_\text{penalty}$ can be used to punish gauge-violating excitations, forcing the system to remain near the gauge-invariant subspace if the penalty is large compared to the physical energy scales. Although this approach can be easier to implement on hardware without built-in symmetry, it relies on a finite energy gap, so gauge violations may persist at some small amplitude.
%
%
	
	\item Packaged States:  
	In contrast to the above methods, packaged states ``hardwire'' gauge invariance by restricting each site/link to an appropriate irrep space. This obviates the need for large penalty energies or a gauge choice but may require more complex qudit or spin-$S$ hardware encodings \cite{Sala2018}.
	The trade-offs include:	
	\begin{itemize}
		\item \textbf{Simplicity vs.\ Overhead:} 
		Penalty-based methods allow using more conventional qubit mappings with no built-in symmetry, but require carefully tuned $\lambda$ and additional gates to implement $\hat{H}_\text{penalty}$.
		Packaged states require building gauge‐irrep subspaces for every link/site from scratch, which can raise the local dimension or complexity of the encoding, yet eliminates gauge violations entirely.
		
		\item \textbf{Robustness to Gauge‐Violating Errors:}
		In packaged approaches, operators that would violate Gauss’s law simply act as zero or move the state out of the physical code space in a detectable manner.
		In penalty‐term approaches, small errors might excite gauge‐violating states if $\lambda$ is not sufficiently large.
		
		\item \textbf{Flexibility for Mixed Strategies:}
		One can combine partial gauge‐fixing or penalty terms with a partially packaged representation if that suits a given hardware constraint or if partial constraints are easier to enforce.
	\end{itemize}
\end{enumerate}

\begin{table}[H]
	\centering
	\caption{Comparison of methods for enforcing gauge invariance in quantum simulation of lattice gauge theories.}
	\begin{tabular}{|p{2cm}|p{6.5cm}|p{6.5cm}|}
		\hline
		\textbf{Method} & \textbf{Pros} & \textbf{Cons} \\
		\hline
		\textbf{Packaged States} &
		\begin{itemize}
			\item Enforces gauge invariance exactly by construction.
			\item Automatically excludes unphysical (gauge-violating) states.
			\item Naturally encodes full irreps (no partial charges or flux).
		\end{itemize} 
		&
		\begin{itemize}
			\item May require higher local Hilbert-space dimensions or more complex encodings.
			\item Implementation on hardware might be challenging if available resources are limited.
		\end{itemize} \\
		\hline
		\textbf{Gauge Fixing} &
		\begin{itemize}
			\item Reduces redundant degrees of freedom by choosing a specific gauge.
			\item Simplifies the Hilbert space by eliminating some gauge-variant modes.
		\end{itemize} 
		&
		\begin{itemize}
			\item May lead to residual gauge ambiguities or global constraints.
			\item Can introduce complications with boundary conditions or Gribov ambiguities in non-Abelian theories.
		\end{itemize} \\
		\hline
		\textbf{Penalty Terms} &
		\begin{itemize}
			\item Can be implemented on hardware that does not natively support gauge-invariant encodings.
			\item Flexible by tuning the penalty strength to suppress gauge violations.
		\end{itemize} 
		&
		\begin{itemize}
			\item Finite penalty strength means that gauge-violating errors are not completely eliminated.
			\item Requires careful tuning relative to physical energy scales.
			\item Leakage into unphysical sectors may accumulate over long simulation times.
		\end{itemize} \\
		\hline
	\end{tabular}
	\label{tab:comparison_methods}
\end{table}

Overall, packaged quantum states present a conceptually clear and computationally promising framework for simulating strongly coupled gauge theories on near-term quantum devices. The choice between packaged states, gauge fixing, or penalty-term methods will ultimately depend on the specifics of the hardware platform, the target model, and the desired precision in suppressing gauge violations. We anticipate that integrating these approaches with robust error-correction and noise mitigation techniques will be key to unlocking the full potential of quantum simulations in high-energy physics and beyond.

\section{Conclusion}

We have introduced an approach to quantum simulation of lattice gauge theories based on packaged quantum states, which encode each matter or gauge excitation in a complete irreducible representation of the local gauge group. By requiring that every local excitation carry a full (rather than partial) charge or color, this construction automatically enforces Gauss’s law and superselection rules, eliminating the need for large penalty terms or gauge-fixing procedures to suppress unphysical states.

Although single-particle packaged states exhibit no multi-particle packaged entanglement, multi-particle superpositions confined to one gauge sector can form nontrivial packaged entangled states that encode color-singlet hadrons, flux loops, and other gauge-invariant excitations. Beyond these theoretical advantages, the packaging principle also holds practical promise - potentially leading to more efficient circuits or lower gate overhead for simulating lattice gauge theories.

We illustrated this approach with examples from $\mathrm{U}(1)$, $\mathrm{SU}(2)$, and $\mathrm{SU}(3)$ gauge theories. In each case, packaging unifies the description of single-particle excitations and multi-particle packaged entangled states by enforcing local gauge invariance at the operator level. As a result, any gauge-violating process that would generate fractional charges immediately places the system outside the physical subspace, rendering such errors detectable as leakage.

While packaging alone does not correct gauge-respecting errors, it provides a robust foundation for integrating gauge invariance with standard quantum error-correction. As quantum hardware continues to scale, we expect these techniques to enable more realistic simulations of nonperturbative gauge phenomena. Ultimately, packaged states offer both a practical tool for preserving local constraints in near-term devices and a conceptual framework for exploring color confinement and other strongly coupled physics. We anticipate that combining the packaging framework with established quantum-information strategies for error correction and noise mitigation will be crucial for robust simulations of QCD-like theories and beyond.

\end{document}